\documentclass[11pt,a4paper]{article}

\usepackage{multirow}
\usepackage{amsmath}
\usepackage{mathrsfs}
\usepackage{bbm}
\usepackage{amsfonts}
\usepackage{graphicx}
\usepackage{subfig}
\usepackage{setspace,epstopdf,amsfonts,amssymb,amsthm}
\usepackage{marginnote,enumitem,rotating,fancyvrb}
\usepackage{inputenc}
\usepackage[hidelinks]{hyperref}
\usepackage{geometry}
\usepackage{setspace} 
\usepackage{tikz} 
\usetikzlibrary{decorations}
\usepackage{hyperref,float}
\usepackage{exceltex}
\usepackage{booktabs}
\usepackage{tabulary}
\usepackage{adjustbox} 
\usepackage{accents}
\usepackage{comment}
\usepackage{braket}

\usepackage[round]{natbib}   
\newtheorem{corollary}{Corollary}

\newtheorem{definition}{Definition}

\newtheorem{lemma}{Lemma}

\newtheorem{proposition}{Proposition}

\newcommand{\eg}{\emph{e.g.,}}

\hypersetup{
	colorlinks=true,
	linkcolor=red,
	citecolor=blue
}

\title{A Search Model of Statistical Discrimination\thanks{We thank Simon Alder, Luca Flabbi, Teddy Kim, Fei Li, Anja Prummer, and Huseyin Yildrim for comments and discussions. The usual disclaimer applies.} }

\author{Jiadong Gu\thanks{Department of Economics, University of North Carolina at Chapel Hill, 107
		Gardner Hall, CB 3305, Chapel Hill, NC 27599-3305. Email: \texttt{jiadong@live.unc.edu}} \and Peter Norman\thanks{
		Department of Economics, University of North Carolina at Chapel Hill, 107
		Gardner Hall, CB 3305, Chapel Hill, NC 27599-3305. Email: \texttt{normanp@email.unc.edu}}}

\date{\today}

\begin{document}

\maketitle

\begin{abstract}

We offer a search-theoretic model of statistical discrimination, in which firms treat identical groups unequally based on their occupational choices. The model admits symmetric equilibria in which the group characteristic is ignored, but also asymmetric equilibria in which a group is statistically discriminated against, even when symmetric equilibria are unique. Moreover, a robust possibility is that symmetric equilibria become unstable when the group characteristic is introduced. Unlike most previous literature, our model can justify affirmative action since it eliminates asymmetric equilibria without distorting incentives.
	
	\smallskip
	
	
	\thispagestyle{empty}
	\newpage 
\end{abstract}

\pagenumbering{arabic}

\section{Introduction}
\textit{Statistical discrimination} refers to situations in which some agents use \textit{observable characteristics} as a proxy for payoff relevant \textit{unobservable characteristics}. The observable characteristics on which statistical discrimination may be based include gender, race, job separation rate, unemployment duration, job leave duration, or anything else that may serve as a proxy for unobservables that the market cares about. Models of statistical discrimination were initially developed as an alternative to taste-based models to explain group inequalities (see \citet{fang2011theories}). There are many variants, but the main point with many of these models is that interactions between a signal extraction problem and human capital investments can generate equilibria in which some group(s) are worse off than others despite no fundamental differences between groups (see \citet{arrow1973theory}, \citet{coate1993will}, and \citet{moro2004general}).\footnote{The first paper along these lines is \citet{arrow1973theory}. See also \citet{coate1993will}, \citet{moro2004general}, and the survey by \citet{fang2011theories}.}

We consider statistical discrimination in a frictional search framework. This allows us to explore an alternative channel in which self-fulfilling statistical discrimination can be sustained that is complementary to the existing literature. Instead of investments in human capital, inequalities arise in equilibrium due to search frictions and occupational choice. As demonstrated by \citet{xiao2020wage}, differences in occupational choice account for much of the gender wage gap, but it is accounted for as reflecting preferences for amenities. This paper provides a complementary explanation, as it shows that women may enter low paying occupations as a result of statistical discrimination.

Our baseline model without group characteristics is a discrete time random search model in which there are two types of workers and two distinct technologies. We label the workers as \textit{qualified} or \textit{unqualified} and the technologies as \textit{high tech} and \textit{low tech. }Conditional on an appropriate match, a high tech firm is more productive than a low tech firm. However, only qualified workers are productive at high tech firms, whereas the type of worker does not matter for low tech firms. This type of technology has been considered before in the labor-search literature by \citet{albrecht2002matching}, \citet{gautier2002unemployment}, \citet{dolado2009job}, and others. However, as far as we know, the only paper that has combined a technology along these lines with asymmetric information about worker type is \citet{jarosch2018statistical}, but they have an equilibrium characterization very different from ours.

When a high tech firm matches with a worker, they would like to hire the worker if and only if the worker is qualified. Because the firm
cannot directly observe the type of worker, this ideal hiring rule can not be implemented. Instead, the firm observes a noisy signal that is correlated with the type and may be interpreted as the result of a job interview. The signal is labeled so that a higher signal is good news, which implies that an optimal hiring rule is one in which a worker is offered a job if and only if the signal exceeds some threshold. 

Importantly, the optimal hiring threshold depends not only on parameters of the noisy signal but also on the \textit{prior probability that a worker is qualified}. Since we consider random matching, this prior probability is simply the proportion of qualified workers in the pool of unemployed, which is endogenous. Hence, while search externalities along the lines of many existing papers are present in our model, there are now also \textit{informational externalities} that are novel to the search and matching literature. 

If the probability that a qualified worker accepts a low tech position increases, then the outflow of qualified workers from the pool of unemployed increases. In steady state, this reduces the proportion of qualified workers in the pool of unemployed, which increases the hiring threshold. Hence, a steady state increase in the probability that a qualified worker accepts a low tech job makes it harder to obtain a high tech job. With free entry, this also affects entry decisions and the details of how that works are somewhat intricate and also depend on whether workers are following a pure or mixed acceptance rule. The main point, however, is that there is feedback from the worker acceptance rule in the low tech sector to the optimal hiring threshold for high tech firms.

The baseline model may have a unique equilibrium, or there may be multiple equilibria. However, it is important to note that nothing in our analysis with multiple groups rests on multiplicity in the single group model. We introduce a payoff irrelevant observable characteristic by assuming that a worker either is from group $m$ or group $f$, which can be observed by the firms. The two groups are identical in the sense that the proportion of qualified workers is the same for both groups, but we allow the groups to be of different sizes. We demonstrate that the existence of asymmetric equilibria is a robust possibility, whether or not the baseline model has a unique equilibrium.

If everyone ignores the observable characteristic, the equilibrium conditions in the case with multiple groups are the same as in the baseline
model, so there is always at least one equilibrium with equal treatment of the workers. However, these symmetric equilibria may be very fragile in the case of multiple groups. One possible symmetric equilibrium is when both sectors are active, and qualified workers randomize between accepting low tech jobs. In such an equilibrium, firms are indifferent between entering as a high tech and as a low tech firm, and the proportion of high tech firms is determined so that qualified workers are indifferent between accepting and rejecting low tech job offers. Imagine that the proportion of qualified $m$ workers increases ever so slightly while the proportion of qualified $f$ workers is held fix. Assuming no change in the fraction of high tech jobs, this would make the best response for $m$ workers to reject low tech jobs for
sure. In the full equilibrium, the proportion of high tech job will in general change, but this does not change the argument much. Either $m$
workers have to reject low tech jobs for sure or $f$ workers have to accept low tech jobs for sure after the change. Therefore, an arbitrarily small exogenous change in the proportion of qualified men creates a significant difference through the equilibrium conditions. Hence, symmetric equilibria of the form just described can not be stable.

A related point is that there is a true interaction between groups. Unlike models like \citet{coate1993will}, in which discrimination is interpreted as one group coordinating on a good equilibrium and another on a bad one, incentives for one group are affected by the behavior of the other. This non-separability allows asymmetric equilibria to exist even if the baseline model has a unique equilibrium.

Most existing models of statistical discrimination focuses on the interplay between incentives for human capital investments and hiring decisions. A
group is discriminated in the labor market because, \textit{in equilibrium,} the group is less skilled on average. In our frictional model of the labor market, the average skill level \textit{in the pool of unemployed} is not the same as the skill level in the population, so we are able to explain statistical discrimination between groups that in equilibrium are equally skilled. 

The standard model explains the lower skill level as the consequence of having less high powered incentives to invest in skills. It has been argued that there is very little empirical evidence for this. In fact, in the case of the black-white wage gap, \citet{neal1996role} and \citet{neal2006has} argue that blacks have stronger incentives to acquire skills than whites, which is inconsistent with standard models of statistical discrimination.\footnote{However, \citet{glover2017discrimination} provides some evidence in favor of models of statistical discrimination.} In our model, skills are exogenous, but statistical discrimination is still a possibility. Instead of feeding
into incentives for skill acquisition, the labor market responses feed into incentives to accept dead-end jobs. So, in a sense, the decision to turn down bad jobs in our model plays a similar role as skill investments in the standard model, and there are similar free-riding considerations involved as individuals benefit from the total number of people within the group that turn low tech jobs down.

Another issue with standard models of statistical discrimination is that, in equilibrium, the discriminated group is, on average, less productive than the dominant group. This is possible despite there being no intrinsic differences between groups, but a theory that implies that women are significantly less productive than men because of the lack of human capital investments may not be the most plausible. After all, women are now acquiring more education than men. While education is not the kind of unobservable investments that are considered in models of statistical discrimination, we find our alternative explanation quite plausible. There is also rather convincing evidence of various forms of mismatch between worker skills and jobs (for example, see \citet{clark2017career} and the references therein on worker over-education) suggesting that it seems reasonable to have mismatch also with respect to unobservable skills.

Affirmative action policies can be counterproductive in many conventional models of statistical discrimination. In \citet{coate1993will}, the problem is that preferential treatment may reduce the incentives to acquire skills. This may also be true in \citet{moro2003affirmative}, where, additionally, the targeted group may be may worse off, and such perverse welfare effects are even more prevalent in \citet{fang2006government}. In contrast, a hiring quota requiring firms to hire workers in accordance with population proportions in this model eliminates all asymmetric equilibria. This is because firms must have lower standards for women if there are fewer qualified women in the pool of unemployed. This unambiguously makes qualified women having a better chance than men at high tech jobs, which moves incentives in the desired direction. In contrast, in the previous literature, the hiring quota may create disincentives to invest in human capital.    

We make many simplifying assumptions in order to make our analysis as transparent as possible. Most notably, we assume that wages are exogenous.
While this is obviously unrealistic, we see no reason why endogenizing wages through posting or bargaining would qualitatively change anything.

Our model is not the first dynamic model of statistical discrimination. However, existing dynamic models are very different from our setup. An early contribution is \citet{mailath2000endogenous}, in which firms can direct their search towards specific groups, creating discriminatory equilibria. In \citet{jarosch2018statistical} some elements are similar, but they have a unique equilibrium and consider \textquotedblleft statistical
discrimination\textquotedblright\ with respect to unemployment duration, which is a proxy for skills because a long unemployment spell signals that the worker has been rejected by many firms. In \citet{kim2018collective} it is shown that, depending on parameters, it may or may not be possible to escape an undesirable equilibrium by re-coordinating beliefs in a stylized overlapping generations version of \citet{coate1993will}. In \citet{bohren2019dynamics} beliefs evolve in equilibrium, but the underlying source of discrimination is irrational biased beliefs. Our model does not need such exogenous bias. Finally, \citet{che2019statistical} consider a model of ratings-guided markets which is very different from ours in terms of modelling details, but in which similar externalities arise. Like in our model, discriminatory equilibria may also arise in the case in which there is a unique symmetric equilibrium.\footnote{ See also \citet{antonovics2006statistical}, \citet{eeckhout2006}, \citet{glawtschew2015eliminating}, and  \citet{masters2014statistical} }

\section{The Baseline Model}
We consider a labor market where a unit mass of workers and a continuum of firms are matched randomly. Workers and firms are infinitely lived, forward-looking, risk-neutral, and have a common discount factor of $\beta$. Time is discrete, and we focus on steady state equilibria.

Later on, we will add observable payoff irrelevant group characteristics that can be interpreted as race or gender, but to minimize notation we begin by introducing a benchmark model with no group characteristics. This baseline model also provides a full characterization of \textit{symmetric equilibria} in the model with observable group characteristics.

A proportion $\psi $ of the workers are \textit{qualified }, and the remaining fraction $1-\psi $ are referred to as \textit{unqualified}. 
Qualified workers are equipped with a skill that matters for some firms, but not for others. Specifically, we assume that a firm can be one of two types. Some firms, referred to as \textit{low tech }firms, produce flow output $y_{l}$ and pays exogenous wage $w_{l}$ when matched with a worker of either type. We assume that $y_{l}-w_{l}>0$ implying that low tech firms are willing to employ any worker it matches with.

Other firms face a non-trivial decision when matched with a worker. If a \textit{high tech }firm is matched with a qualified worker the match
produces flow output $y_{h}$ and the firm pays the exogenous wage $w_{h}$, where $y_{h}-w_{h}>0$. The firm would thus want to hire the worker if the worker is known to be qualified. In contrast, if the worker is unqualified the flow output is (normalized to) zero, so the flow profit is $-w_{h}$, so a high tech firm would never want to hire an unqualified worker. We assume that $w_{l}<w_{h}$, implying that high tech matches are more desirable than low tech matches for the worker. The flow value of unemployment is $b<w_{l}$.

The firm cannot directly observe whether a worker is qualified or not. Instead, the firm observes a noisy signal $\theta \in \lbrack 0,1]$, which is drawn from density $f_{q}(\theta )$ if the worker is qualified and from $f_{u}(\theta )$ otherwise. Since $\text{the signal }\theta $ is pure information, there is no loss of generality to label the signal realizations so that higher realizations correspond with a higher probability of the worker being qualified, so we assume that the likelihood ratio $\frac{f_{q}(\theta )}{f_{u}(\theta )}$ is strictly increasing in $\theta $.\footnote{Assuming that the likelihood ratio is \textit{strictly} increasing and assuming away mass points at the same time is not completely without loss. We make the assumption to assure a unique solution for the firm optimal hiring strategy. Weakening to weak monotonicity allows for signals replicating discrete support. This is less elegant, but can be dealt with.} 

Let $\pi $ be the firm's prior probability that a worker is qualified. Using the monotone likelihood ratio property, it is immediate that the posterior
probability after observing $\theta $
\begin{equation}
P\left( \theta ,\pi \right) =\frac{\pi f_{q}\left( \theta \right) }{\pi f_{q}\left( \theta \right) +\left( 1-\pi \right) f_{u}\left( \theta \right) },  \label{bayesrule}
\end{equation}
is also monotone in $\theta ,$, which makes it very easy to describe optimal hiring rules.

If a match forms we also assume that the qualification of the worker is revealed to the corresponding matched firm in every period with some
probability $r$. Should a worker in a high skilled firm be revealed to have low productivity, that worker is fired. There is also an exogenous separation probability $\phi$, which is ensuring that steady state conditions behave nicely. 

We parametrize the model so that the unqualified workers' acceptance rules are trivial: unqualified workers accept any job offer they get. In contrast, a qualified worker will always accept a high tech offer as $b<w_{l}<w_{h}$, but may or may not accept low tech jobs depending on, for example, how likely it is to match with a high tech firm in the future. This is the only non-trivial worker choice in the model, and we let $\alpha \in \left[ 0,1\right] $ denote the endogenous probability that a qualified worker accepts when offered a position at a low tech firm.

There is free entry into both the high and the low tech sectors, and the flow cost of a vacancy is $K>0$ in either sector. For simplicity we assume a matching protocol in which the short side gets served and the long side is rationed. Because entry is on the firm side only we never need to worry that there are more workers than firms if the entry cost is set so that there is a strict incentive to enter if the firm matches for sure. Hence, workers will match for sure in each period, but the probability of matching with high tech or low tech firms is endogenously determined by firm entry as the ratio of workers to firms.

To sum up, the only non-trivial choice variables in the model are:
\begin{enumerate}
	\item[(1)] The probability $\alpha \in \left[ 0,1\right] $ for a qualified worker to accept low tech jobs.
	\item[(2)] Entry decisions by firms in both sectors.	
	\item[(3)] Given the noisy signal $\theta \in \left[ 0,1\right]$, whether a high tech firm should offer a job to the worker in a match.
\end{enumerate}

We will focus solely on steady state equilibria.

\section{ Equilibria in the Baseline Model}

\subsection{Firms' Problem}
We begin by considering the hiring decision for a \textit{high-tech firm}. Let $\pi \in \left[ 0,1\right] $ be the endogenous stationary proportion of qualified workers in the pool of unemployed, which is also the prior probability that the worker is qualified from the point of view of the firm. Also, let $W\left( 0,\pi \right) ,W\left( u,\pi \right) $ and $W\left(q,\pi \right) $ be the value of a vacancy, being matched with an unqualified worker, and being matched with a qualified worker respectively, as a function of $\pi $. The firm values of hiring an unqualified and qualified worker are
\begin{eqnarray}
W\left( u,\pi \right) &=& -w_{h}+\beta \left[ \left( \phi +\left( 1-\phi\right) r\right) W\left( 0,\pi \right) +\left( 1-\phi \right) \left(1-r\right) W\left( u,\pi \right) \right]  \\
W\left( q,\pi \right) &=& y_{h}-w_{h}+\beta \left[ \phi W\left( 0,\pi \right) +\left( 1-\phi \right) W\left( q,\pi \right) \right]. \notag
\end{eqnarray}
For the unqualified worker, the probability of moving from employment at a high tech firm to unemployment is higher than for the qualified worker. This is because the type may be revealed in addition to the common separation rate. Solving for these values in terms of the value of a vacancy and using the free entry condition $W\left( 0,\pi \right)=0$, we can express these values purely in terms of the exogenous parameters as
\begin{eqnarray}
W\left( u,\pi \right)  &=&\frac{-w_{h}}{1-\beta \left( 1-\phi \right) \left(1-r\right) }\equiv W_{u}<0  \label{constantfirmvalues} \\
W\left( q,\pi \right)  &=&\frac{y_{h}-w_{h} }{1-\beta \left( 1-\phi \right) }\equiv W_{q}>0.  \notag
\end{eqnarray} 
Now, assume that a firm has matched with a worker and that the noisy signal $\theta \in \left[ 0,1\right] $ has been drawn. If the worker is hired, the expected continuation payoff is $W_{q}$ if the worker is qualified and $W_{u}$ if the worker is not. In contrast, if the worker is not hired, the firm moves on to the next period with a vacancy, which has value $0.$ Hence, the
firm is better off hiring the worker in expectation if and only if
\begin{equation}
P\left( \theta ,\pi \right) W_{q}+\left( 1-P\left( \theta ,\pi \right) \right) W_{u}\geq 0,  \label{hiringprofit}
\end{equation}
where $P\left( \theta ,\pi \right)$ is defined in~(\ref{bayesrule}). Since $P\left( \theta ,\pi \right) $ is strictly increasing in $\theta$ for any $\pi \in \left[ 0,1\right] $ we have that:

\begin{lemma}\label{cutoff}
For any given $\pi \in \left[ 0,1\right] $ the unique optimal hiring rule is a threshold rule where a worker is offered a job if and only
$\theta \geq s\left( \pi \right).$ If $\frac{\pi f_{q}\left( 0\right) }{\left( 1-\pi\right) f_{u}\left( 0\right) }\geq -\frac{W_{u}}{W_{q}}$ then $s\left( \pi \right) =0$ and $\frac{\pi f_{q}\left( 1\right) }{\left( 1-\pi \right) f_{u}\left( 1\right) }\leq -\frac{W_{u}}{W_{q}}$ then $s\left( 1\right) =0.$ If $\frac{\pi f_{q}\left( 0\right) }{\left( 1-\pi \right) f_{u}\left( 0\right) }<-\frac{W_{u}}{W_{q}}<\frac{\pi f_{q}\left( 1\right) }{\left(1-\pi \right) f_{u}\left( 1\right) }$ the threshold $s\left( \pi \right) \in \left( 0,1\right)$ satisfies
	\begin{equation}
	\frac{\pi f_{q}\left( s\left( \pi \right) \right) }{\left( 1-\pi \right)f_{u}\left( s\left( \pi \right) \right) }=-\frac{W_{u}}{W_{q}}.
	\label{standard}
	\end{equation}
\end{lemma}

Notice that one can also understand the threshold $s\left( \pi \right) $ as the solution to 
\begin{equation}
\max_{\theta }\pi \left[ 1-F_{q}(\theta )\right] W_{q}+(1-\pi )\left[1-F_{u}(\theta )\right] W_{u}.  \label{maxproblem}
\end{equation}
The interpretation of this problem is as the ex ante expected profit from a threshold rule. This is because $\pi $ is the probability that a worker is qualified and $1-F_{q}\left( \theta \right) $ is the conditional probability that the signal is above $\theta ,$ so $\pi \left( 1-F_{q}\left( \theta \right) \right) $ is the probability of hiring a qualified worker if the hiring threshold is $\theta .$ Symmetrically, $\left( 1-\pi \right) \left(1-F_{u}\left( \theta \right) \right) $ is the probability of hiring an unqualified worker if the hiring threshold is $\theta $, so the objective function in (\ref{maxproblem}) is the profit as a function of the hiring threshold. The condition (\ref{standard}) can be obtained from the first order condition for the problem (\ref{maxproblem}).

For a \textit{low tech firm}, it always offers a job to any worker following a match. The continuation payoff from hiring is
\begin{equation}\label{wl}
W_l=\frac{y_l-w_l}{ 1-\beta (1-\phi) } >0,
\end{equation}
using the free entry condition as for high tech firms.

\subsection{Equilibrium Entry}
For notational simplicity let
\begin{eqnarray}\label{Adef}
	A_{q}\left( \pi \right)  &=&1-F_{q}\left( s\left( \pi \right) \right)  \\
	A_{u}\left( \pi \right)  &=&1-F_{u}\left( s\left( \pi \right) \right) \notag
\end{eqnarray}
be the unique probabilities that qualified and unqualified workers are hired when matched with a high tech firm in an equilibrium in which the proportion of qualified workers in the pool of unemployed in $\pi .$ It should be intuitively clear that these probabilities will affect the incentives for qualified workers. We therefore write $\alpha \left( \pi \right)$ for the probability that a qualified worker accepts a low tech job explicitly as a function of $\pi $ despite not yet having discussed the worker optimization problem. The free entry conditions for high each and low tech sectors are then 
\begin{eqnarray}
0 &=&-K+\beta p_{f}\left[ \pi A_{q}(\pi )W_{q}+(1-\pi )A_{u}(\pi )W_{u}\right]   \label{fe1} \\
0 &=&-K+\beta p_{f}\left[ \pi \alpha \left( \pi \right) +(1-\pi )\right] W_{l},  \notag
\end{eqnarray}
where $p_{f}$ is the probability that a firm matches with a worker, which is equal to the ratio of
unemployment over vacancies, or the inverse market tightness rate. Note that

\begin{lemma}\label{profit}
	The profit of entering high tech sector, $ \pi A_{q}(\pi) W_{q} + (1-\pi) A_{u}(\pi) W_{u}$, is strictly increasing in $\pi$.
\end{lemma}

The intuitive idea is that the profit is strictly increasing in $\pi$ for any fixed hiring threshold, and adjusting the threshold only increases the gain. The proof is in the appendix. It follows that $p_{f}$ is strictly decreasing in $\pi$, given that the high tech sector is active. Hence, total entry is strictly increasing in $\pi,$ except in the uninteresting case in which no high tech firms enter.

However, a condition that will prove more useful for the equilibrium characterization is that (\ref{fe1}) implies that when both sectors are
active, firms must be indifferent between entering as high or low tech firms, or 
\begin{equation} \label{freeentry}
\left[ \pi \alpha \left( \pi \right) +(1-\pi )\right] W_{l}=\pi A_{q}(\pi)W_{q}+(1-\pi )A_{u}(\pi )W_{u}.  
\end{equation}
Notice that the assumption that $K$ is equal for the two sectors is just a normalization, as differences in costs of maintaining a vacancy can be incorporated in $W_{l}.$ Next, consider the possibility that qualified workers always reject low tech wage offers. In order for both sectors to be active in this case, it must be that the proportion of qualified workers among the unemployed is $\underline{\pi }$, which we define as the unique solution to 
\begin{equation} \label{picky}
(1-\underline{\pi })W_{l}=\underline{\pi }A_{q}(\underline{\pi })W_{q}+(1-\underline{\pi })A_{u}(\underline{\pi })W_{u},
\end{equation}
The expected profit of being in the low tech sector decreases in $\pi $ and the expected profit of being in the high tech sector increases in $\pi$, so if $\pi <\underline{\pi }$ then only low tech firms would be willing to enter. Symmetrically, if qualified workers always accept low tech jobs the proportion of qualified workers that ensure indifference, $\overline{\pi }$,
is 
\begin{equation}\label{notpicky}
W_{l}=\overline{\pi }A_{q}(\overline{\pi })W_{q}+(1-\overline{\pi })A_{u}(\overline{\pi })W_{u}.
\end{equation}

Hence, if $\pi > \underline{\pi }$ the expected payoff of being a high tech firm is higher than being a low tech firm even if qualified workers always accept low tech offers. Summing this up we have that:

\begin{lemma}\label{firmindifference}
	Suppose that $K$ is small enough so that there are firms that want to enter. Then
\begin{enumerate}[label=(\arabic*)]
	\item If $\pi <\underline{\pi }$ then firms will enter the low tech sector only regardless of what workers do.
	\item If $\pi =\underline{\pi }$ then firms are willing to enter in each sector if and only if $\alpha \left( \underline{\pi }\right) =0$. If $\alpha \left( \underline{\pi }\right) >0$ firms are willing to enter the low tech sector only.
	\item If $\underline{\pi }<\pi <\overline{\pi }$ there is a unique $\alpha\left( \pi \right) \in (0,1)$ such that firms are willing to enter in each sector. For $\alpha>\alpha\left( \pi \right) \in (0,1)$ only low tech firms are willing to enter and if $\alpha<\alpha\left( \pi \right) \in (0,1)$ only high tech firms are willing to enter.
	\item If $\pi =\overline{\pi }$ firms are willing to enter in each sector if and only if $\alpha \left( \overline{\pi }\right) =1$. If $\alpha \left( \overline{\pi }\right) <1$ only high tech firms are willing to enter.
	\item If $\pi >\overline{\pi }$ firms will enter the high tech sector only regardless of what workers do.
\end{enumerate}
\end{lemma}
The firms' optimality is summarized into the following Figure \ref{fig_firmsoptimality}.  The details of the proof are in Appendix~\ref{app_firmindifference}.
	\begin{figure}[H]
	\centering
	\begin{tikzpicture}[xscale=.85]
	\draw (0,0) -> (15,0);
	\draw (0,-.2) -- (0, .2);
	\draw (5,-.2) -- (5, .2);
	\draw (10,-.2) -- (10, .2);
	\draw (15,-.2) -- (15, .2);
	
	\draw (0,-.2) node [below] {$0$} (5,-.2) node [below] {$\underline{\pi}$} (10,-.2) node [below] {$\overline{\pi}$} (15,-.2) node [below] {$1$};
	
	\node[align=left, below]   at (2.5,1) {Low tech};
	\node[align=center, below] at (7.5,1) {Two sectors};
	\node[align=right, below]  at (12.5,1) {High-tech};
	\end{tikzpicture}
	\caption{Firm Entry as Function of Beliefs.}
	\label{fig_firmsoptimality}
\end{figure}
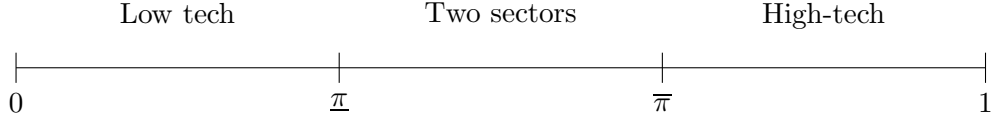

\subsection{Workers' Optimization}

Since we have trivialized the unqualified workers into agents who always accept every offer, we only need to consider the qualified workers. Denote by $V_{0}\left( \pi \right)$, $V_{h}\left( \pi \right) $ and $V_{l}\left( \pi \right) $ the value for a qualified worker from being unemployed, being employed in the high tech sector, and being employed in the low tech sector, respectively. Also, let $p\left( \pi \right) $ be the endogenous probability that the worker meets a high tech firm, so that $1-p\left( \pi \right) $ is the probability of meeting a low tech firm. We use the convention that $V_{0}\left( \pi \right) $ is the value of unemployment immediately prior to matching with the firm, so that
\begin{eqnarray}
V_{0}\left( \pi \right) = &p\left( \pi \right)& \left[ A_{q}\left( \pi\right) V_{h}\left( \pi \right) +\left( 1-A_{q}\left( \pi \right) \right)\left( b+V_{0}\left( \pi \right) \right) \right]   \label{uvalue} \\
+ &\left( 1-p\left( \pi \right) \right)& \max_{\alpha \in \left[ 0,1\right] } \left[ \alpha V_{l}\left( \pi \right) +\left( 1-\alpha \right) \left[ b+\beta V_{0}\left( \pi \right) \right] \right] ,  \notag
\end{eqnarray}
where we can write the values of being employed in terms of exogenous parameters and the value of unemployment as%
\begin{eqnarray}
V_{h}\left( \pi \right)&=&\frac{w_{h}+\phi \beta V_{0}\left( \pi \right) }{1-\left( 1-\phi \right) \beta }  \label{handlvalues} \\
V_{l}\left( \pi \right)&=&\frac{w_{l}+\phi \beta V_{0}\left( \pi \right) }{1-\left( 1-\phi \right) \beta }.  \notag
\end{eqnarray}%
Hence, letting $\alpha \left( \pi \right) $ be the solution to the optimization problem in (\ref{uvalue}). Then,%
\begin{equation}
\alpha \left( \pi \right) =\left\{ 
\begin{array}{cc}
1&\text{if }\frac{w_{l}+\phi \beta V_{0}\left( \pi \right) }{1-\left(1-\phi \right) \beta }>b+\beta V_{0}\left( \pi \right)  \\ 
\left[ 0,1\right]&\text{if }\frac{w_{l}+\phi \beta V_{0}\left( \pi\right) }{1-\left( 1-\phi \right) \beta }=b+\beta V_{0}\left( \pi \right) \\ 
0 & \text{if }\frac{w_{l}+\phi \beta V_{0}\left( \pi \right) }{1-\left(1-\phi \right) \beta }<b+\beta V_{0}\left( \pi \right) 
\end{array}
\right..   \label{acceptreject}
\end{equation}%
Let $V^{\ast }$ be the value of unemployment that makes the worker indifferent between accepting and rejecting, that is%
\begin{equation} \label{Vstar1}
\frac{w_{l}+\phi \beta V^{\ast }}{1-\left( 1-\phi \right) \beta }=b+\beta V^{\ast }.
\end{equation}%
Then, since the value of working in the low tech sector increases in the value of unemployment at a rate strictly slower than $\beta$ we can express (\ref{acceptreject}) as  
\begin{equation}
\alpha \left( \pi \right) =\left\{ 
\begin{array}{cc}
1 & \text{if }V_{0}\left( \pi \right) <V^{\ast } \\ 
\left[ 0,1\right]  & \text{if }V_{0}\left( \pi \right) =V^{\ast } \\ 
0 & \text{if }V_{0}\left( \pi \right) >V^{\ast }%
\end{array}%
\right.. \label{acceptreject2}
\end{equation}%
Combining with (\ref{uvalue}) we have that for the worker to be indifferent it must be that%
\begin{eqnarray}
V^{\ast } 
&=&p\left( \pi \right) A_{q}\left( \pi \right) \frac{w_{h}+\phi \beta V^{\ast }}{1-\left( 1-\phi \right) \beta }+\left( 1-A_{q}\left( \pi \right) p\left( \pi \right) \right) \left( b+\beta V^{\ast }\right).   \notag
\end{eqnarray}

Let $p\left(\pi\right)A_q\left(\pi\right)=Q\left(\pi\right)$ be the probability of meeting a high-tech firm and being hired for a qualified worker, we have
\begin{equation}
	V^\ast =Q\left( \pi \right) \frac{w_{h}+\phi \beta
	V^{\ast }}{1-\left( 1-\phi \right) \beta }+\left( 1- Q\left( \pi \right) \right) \left( b+\beta V^{\ast }\right).   \label{Q}
\end{equation}

Hence, the probability for a qualified worker to be hired by a high tech firm that makes the worker indifferent between accepting and rejecting low tech jobs can be determined independently from $\pi$. We thus get the following simple characterization of the worker acceptance rule.

\begin{lemma} \label{workeroptimal}
	Suppose that $0 \leq w_l-b \leq \beta (1-\phi) (w_h - b)$. Then there exists a unique constant $Q^{\ast }\in \left[ 0,1\right] $ such that%
	\begin{equation}
	V^{\ast } = Q^{\ast }\frac{w_{h}+\phi \beta V^{\ast }}{1-\left( 1-\phi \right)\beta } +\left( 1-Q^{\ast }\right) \left( b+\beta V^{\ast}\right),
	\end{equation}%
	where $V^{\ast }$ is defined in (\ref{Vstar1}). Moreover,%
	\begin{equation}
	\alpha \left( \pi \right) =\left\{ 
    \begin{array}{cc}
	1                  & \text{if $p(\pi)A_q(\pi) < Q^\ast$} \\ 
	\left[ 0,1\right]  & \text{if $p(\pi)A_q(\pi) = Q^\ast$} \\ 
	0                  & \text{if $p(\pi)A_q(\pi) > Q^\ast$}
	\end{array}%
	\right..   \label{acceptreject3}
	\end{equation}
\end{lemma}
The proof can be found in the appendix. The parameter restrictions are to avoid corner cases, which in terms of the statement of the Lemma would show up as $Q^{\ast }$ being negative or larger than one. First, if $w_{l}<b$, then workers prefer unemployment to working in the low tech sector, so the low tech sector would always be inactive. Secondly, if $w_{l}-b > \beta
(1-\phi )(w_{h}-b)$ one can show that low tech jobs are always accepted even if $p\left( \pi \right) A_{q}\left( \pi \right) =1$.

\subsection{Steady State Conditions}
Let $N$ be the total mass of \textit{unemployed} workers, and $N\pi$ be the mass of \textit{qualified} workers in the \textit{unemployed} workers, $N\left(1-\pi \right)$ be the mass of \textit{unqualified} worker in the \textit{unemployed} workers. Let $E_{k,s}$ be the mass of workers with skill $k\in\{q,u\}$ \textit{employed} in sector $s\in \{h,l\}$. We can then write the steady state conditions equalizing the outflow and inflow of each pool of qualified-high tech, qualified-low tech, unqualified-high tech, unqualified-low tech as
\begin{eqnarray}
N\pi p(\pi) A_q(\pi) &=& E_{q,h} \phi \label{inflowoutflow} \\
N\pi (1-p(\pi)) \alpha(\pi) &=& E_{q,l}\phi \notag \\
N \left(1-\pi \right) p(\pi) A_u(\pi) &=& E_{u,h} (\phi+(1-\phi)r) \notag \\
N \left( 1-\pi \right) (1-p(\pi)) &=& E_{u,l}\phi. \notag
\end{eqnarray}

Moreover, for each type, the total mass of workers must add up to the mass employed in each sector together with those in unemployment.
\begin{eqnarray}
N \pi +E_{q,h}+E_{q,l}&=&\psi \label{q} \\
N (1-\pi) +E_{u,h}+E_{u,l}&=&1-\psi. \notag
\end{eqnarray}

By substituting from Equations~\eqref{inflowoutflow} into Equation~\eqref{q} we obtain
\begin{eqnarray}
N \pi + \frac{N \pi p(\pi) A_q(\pi) }{\phi }+ \frac{N\pi (1-p(\pi)) \alpha(\pi) }{\phi} &=& \psi \\
N (1-\pi)+\frac{ N (1-\pi) p(\pi) A_u(\pi) }{\phi+(1-\phi)r }+\frac{N (1-\pi) (1-p(\pi)) }{\phi } &=& 1-\psi. \notag
\end{eqnarray}
Finally, by eliminating $N$ we can summarize the steady state condition in a single equation
\begin{equation}\label{ss}
1+\frac{p(\pi)A_u(\pi)}{\phi+(1-\phi)r}+\frac{1-p(\pi)}{\phi}=\frac{1-\psi}{1-\pi} \frac{\pi}{\psi} \left[1+\frac{p(\pi)A_q(\pi)}{\phi}+\frac{(1-p(\pi))\alpha(\pi) }{\phi} \right].
\end{equation}
Equation~\eqref{ss} represents the steady state conditions for the labor market, given qualified worker's acceptance rule $\alpha(\pi)$ and high-tech firms' hiring strategy $A_q(\pi)$ and $A_u(\pi)$.

For the analysis that follows it is useful to define real function $G:[0,1]^3\rightarrow \mathbb{R}$,
\begin{equation}\label{Gdef}
G\left(\pi,\alpha,p \right) =1+\frac{p A_u(\pi)}{\phi+(1-\phi)r}+\frac{1-p}{\phi}-\frac{1-\psi}{1-\pi} \frac{\pi}{\psi} \left[1 +\frac{p A_q(\pi)}{\phi}+\frac{(1-p)\alpha }{\phi} \right].
\end{equation}
Using this notation, we can define an equilibrium compactly as:
\begin{definition}
	An equilibrium is a triple $\left(\pi,\alpha\left(\pi\right),p\left(\pi \right) \right)$ such that i) $\alpha \left( \pi \right) $ satisfies worker optimality condition (\ref{acceptreject3}), and $p\left( \pi \right)$ is consistent with optimal entry and; ii) $G\left(\pi,\alpha(\pi),p(\pi)\right)=0$. 
\end{definition}

There is also a choice of an optimal hiring threshold $s(\pi )$, but this is built into the terms $A_{q}(\pi )$ and $A_{u}(\pi )$ in the steady state conditions. Also note that if $0<\alpha \left( \pi \right) <1$, then both sectors must be active implying that any $p\left( \pi \right) \in \left[ 0,1\right] $ is a best response. In this case $p\left( \pi \right) $ are thus set to make the workers indifferent. Finally, note that the total mass of firms entering is also an equilibrium object in principle. However, the matching technology is such that workers match with exactly one firm in every period, so this does not interact with any other variable and can be left implicit in the analysis.

\section{Equilibrium Characterization}
While the steady state equation \eqref{ss}, firm entry, and worker optimality conditions are all rather straightforward, the equilibrating mechanism is different in different ranges. For convenience, we have summarized the various cases in the figure below.
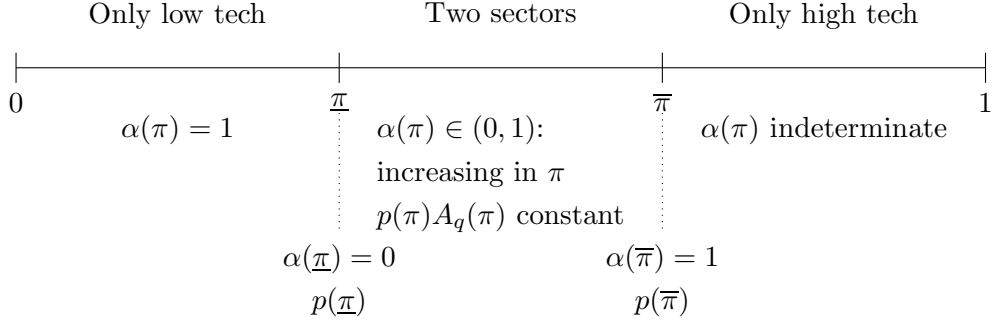
\begin{figure}[H]
	\centering
	\begin{tikzpicture}[xscale=.85]
	\draw (0,0) -- (15,0);
	\draw (0,-.2) -- (0, .2);
	\draw (5,-.2) -- (5, .2);
	\draw (10,-.2) -- (10, .2);
	\draw (15,-.2) -- (15, .2);
	
	\draw [dotted] (5,-.6) -- (5,-2.2); 
	\draw [dotted] (10,-.6) -- (10,-2.2);
	
	\draw (0,-.2) node [below] {$0$} (5,-.2) node [below] {$\underline{\pi}$} (10,-.2) node [below] {$\overline{\pi}$} (15,-.2) node [below] {$1$};
	
	\node[align=left, below]   at (2.5,1) {Only low tech};
	\node[align=center, below] at (7.5,1) {Two sectors};
	\node[align=right, below]  at (12.5,1) {Only high tech};
	
	\node[align=center, below] at (2.5,-.5) {$\alpha(\pi)=1$};
	\node[align=left, below] at (7.5,-.5) {$\alpha(\pi)\in (0,1)$: \\ increasing in $\pi$ \\ $p(\pi)A_q(\pi)$ constant};
	\node[align=center, below] at (12.5,-.5) {$\alpha(\pi)$ indeterminate};
	\node[align=center, below] at (5,-2.2) {$\alpha(\underline{\pi} )=0$\\$p(\underline{\pi})$};
	\node[align=center, below] at (10,-2.2) {$\alpha(\overline{\pi} )=1$\\$p(\overline{\pi})$};
	\end{tikzpicture}
	\caption{Equilibrium Candidates.}
\end{figure}

\subsection{Equilibria with One Sector Active}

Depending on parameters, an equilibrium can have only low tech firms active, only high tech jobs active, or both sectors active. The necessary and sufficient conditions for a low tech equilibrium are straightforward and easy to understand:

\begin{proposition}
	\label{loweq}An equilibrium in which only low tech firms are active exists if and only if $\psi \leq \overline{\pi }.$ This is the unique equilibrium	if $\psi <\underline{\pi }.$
\end{proposition}

The intuition is simple. In a low tech equilibrium, workers become homogenous, so there is no longer any selection. Therefore $\pi =\psi $ in steady state. Recalling that $\overline{\pi }$ defined in (\ref{notpicky}) is the critical value for $\pi$ that makes firms indifferent between the two sectors when all qualified workers accept low tech jobs, we note that for $\pi \leq \overline{\pi }$ firms are more profitable in the low tech sector than in the high tech sector if all workers accept low tech jobs. Moreover, workers have a strict incentive to accept low tech jobs since the probability of a high tech offer is zero. The fact that no other equilibrium can exist when $\pi$ follows directly from the definition of (\ref{picky}).

The conditions for existence of equilibria with only the high tech sector active are a bit more involved for two reasons. Firstly, there will be negative selection effects in such an equilibrium. Secondly, the worker optimality condition is now non-trival. 
We now consider the steady state condition if only high tech firms enter. Since there are no low tech firms around, the determination of $\pi $ in such an equilibrium is independent of $\alpha\left( \pi \right) $ and simplifies to
\begin{equation}
\widetilde{G}\left( \pi \right) =G\left( \pi ,\alpha \left( \pi \right), 1\right) 
=1+\frac{A_{u}(\pi )}{\phi +(1-\phi )r}-\frac{1-\psi }{1-\pi }\frac{\pi }{\psi }\left[ 1+\frac{A_{q}(\pi )}{\phi }\right] =0.
\label{highSS}
\end{equation}%
Note that $\frac{A_{u}(\pi )}{\phi +(1-\phi )r}<\frac{A_{q}(\pi )}{\phi }$ for any $\pi ,$ which reflects that this type of an equilibrium generates negative selection as qualified workers are more likely to move from unemployment to employment and less likely to move from employment to unemployment.

However, while $\alpha \left( \pi \right) $ is irrelevant for the steady state condition we
still have to ask what a worker would do if counter factually matching with a low tech firm. That is, \textit{if} a worker would accept an offer from a deviating low tech entrant and \textit{if }$\pi $ is strictly in between $\underline{\pi }$ and $\overline{\pi }$, then this is inconsistent with a high tech equilibrium. It thus follows:

\begin{proposition}\label{higheq}
	There is at least one solution $\pi ^{\ast }$ to (\ref{highSS}), and;
	
	\begin{enumerate}
		\item Suppose $\pi ^{\ast }<\underline{\pi }$ solves (\ref{highSS}). Then there is no high tech equilibrium corresponding to $\pi ^{\ast }.$
		
		\item Suppose $\pi ^{\ast }\in \left[ \underline{\pi },\overline{\pi }\right]$ solves (\ref{highSS}) and $A_{q}\left( \pi ^{\ast }\right) <Q^{\ast }.$ Then there is no high tech equilibrium corresponding to $\pi ^{\ast }$.
		
		\item Suppose $\pi ^{\ast }\in \left[ \underline{\pi },\overline{\pi }\right]$ solves (\ref{highSS}) and $A_{q}\left( \pi ^{\ast }\right) \geq Q^{\ast }$. Then $( \pi^{\ast }, \alpha\left(\pi^{\ast }\right)=0, p\left(\pi ^{\ast} )=1\right)$ is an equilibrium.
		
		\item Suppose $\pi ^{\ast }>\overline{\pi }$ solves (\ref{highSS}). Then $\left( \pi ^{\ast },\alpha (\pi ^{\ast})=1,p\left( \pi ^{\ast }\right)=1\right) $ is an equilibrium.
	\end{enumerate}
\end{proposition}

\bigskip One may note that $\pi ^{\ast }<\psi $ for any solution to (\ref{highSS}), because if $\pi ^{\ast }$ solves (\ref{highSS}) then
\begin{equation}
1>\frac{1+\frac{A_{u}(\pi ^{\ast })}{\phi +(1-\phi )r}}{1+\frac{A_{q}(\pi^{\ast })}{\phi }}=\frac{1-\psi }{1-\pi ^{\ast }}\frac{\pi ^{\ast }}{\psi },
\end{equation}%
as $\frac{A_{u}(\pi ^{\ast })}{\phi +(1-\phi )r}<\frac{A_{q}(\pi ^{\ast })}{\phi }$ for any $\pi ^{\ast },$. Hence, being unemployed is correlated with being unqualified with being unqualified in an equilibrium with only the high tech sector active. Moreover, $\frac{A_{q}(\pi ^{\ast })}{\phi }\rightarrow \infty $ as $\phi \rightarrow 0$ while $1+\frac{A_{u}(\pi )}{\phi +(1-\phi )r}$ stays finite provided that $r>0.$ Hence, $\pi ^{\ast }\rightarrow 0$ for any sequence of solutions to (\ref{highSS}) as $\phi \rightarrow 0$ implying that there exists $\phi ^{\ast }>0$ such that $\pi ^{\ast }<\underline{\pi }$ for any $\phi \leq \phi ^{\ast }$ and any solution to (\ref{highSS}). We conclude:

\begin{corollary}
	There exists $\phi ^{\ast }>0$ such that no equilibrium in which only high tech firms enter can exist if $\phi \leq \phi ^{\ast }.$
\end{corollary}

\subsection{Equilibria with Both Sectors Active}\label{twosectors}

Having discussed the conditions with the less interesting single sector equilibria, we now consider equilibria in which both sectors are active. Such equilibria may take on different forms in different parts of the parameter space. There are three distinct possibilities:

\begin{enumerate}
	\item All workers accept low tech jobs. This requires that $\pi =\overline{\pi }.$ Additionally, it must be that $p\left( \overline{\pi }\right) A_{q}\left( \overline{\pi }\right) \leq Q^{\ast }$ to justify the decisions of the workers.
	
	\item All workers reject low tech jobs. This requires that $\pi =\underline{\pi }.$ Additionally, it must be that $p\left( \underline{\pi }\right) A_{q}\left( \underline{\pi }\right) \geq Q^{\ast }$ to justify the decisions of the workers.
	
	\item Workers are indifferent and randomize in a way so that firms are indifferent between the two sectors.
\end{enumerate}

\subsubsection{Low Tech Jobs Accepted}\label{twosectors_picky}

First, consider the possibility that $\alpha \left( \pi \right) =1,$ which implies that $\pi =\bar{\pi}$ to ensure firms' indifference. To be consistent with the steady state condition (\ref{ss}) it must be that $p\left( \overline{\pi }\right) $ is such that
\begin{eqnarray}
G\left( \overline{\pi },1,p\left( \overline{\pi }\right) \right) 
= 1+\frac{p\left( \overline{\pi }\right) A_{u}(\overline{\pi })}{\phi +(1-\phi )r}+\frac{1-p\left( \overline{\pi }\right) }{\phi}
-\frac{1-\psi }{1-\overline{\pi }}\frac{\overline{\pi }}{\psi} \left[ 1+\frac{p\left( \overline{\pi }\right) A_{q}(\overline{\pi })}{\phi }+\frac{(1-p\left( \overline{\pi }\right))}{\phi}\right] =0,  \label{notpickySS}
\end{eqnarray}
holds and $p\left( \overline{\pi }\right) A_{q}\left( \overline{\pi }\right)\leq Q^{\ast }.$ In the appendix we show that this is consistent with equilibrium under the following conditions.
\begin{proposition}
	\label{nonpicky}A two sector equilibrium with workers accepting low tech jobs exists if and only if the following conditions hold:
	
	\begin{enumerate}
		\item $\psi >\overline{\pi }$
		
		\item $G\left( \overline{\pi },1,1\right) \leq 0$
		
		\item $p\left( \overline{\pi }\right) A_{q}\left( \overline{\pi }\right) \leq Q^{\ast }$ for the solution to (\ref{notpickySS}).
	\end{enumerate}
	There can be at most one such equilibrium.
\end{proposition}

Notice that the equilibrating variable that balances the steady state condition is no longer $\pi$ for these types of equilibrium. Instead, the steady state is achieved by finding the right mix of low tech and high tech firms. Also note that in this type of equilibrium, the probability of being qualified must be lower for unemployed workers than in the population as a whole, explaining the first condition. Finally, the steady state condition (\ref{nonpicky}) is linear in the proportion of high tech firms, explaining why there can be at most one equilibrium in this form.

\subsubsection{Low Tech Jobs Rejected}\label{twosectors_acc}

Suppose first that both sectors are active and all qualified workers reject low tech jobs. Then $\pi =\underline{\pi }$ and $\alpha \left( \underline{\pi }\right) =0$. We then seek solutions $p\left( \underline{\pi }\right)$ to the steady state equilibrium condition (\ref{ss}) that is consistent with qualified workers rejecting low tech jobs. This is true if and only if%
\begin{eqnarray}
G\left( \underline{\pi },0,p\left( \underline{\pi }\right) \right) 
&=&1+\frac{p\left( \underline{\pi }\right) A_{u}(\underline{\pi })}{\phi +(1-\phi)r}+\frac{1-p\left( \underline{\pi }\right) }{\phi} 
-\frac{1-\psi }{1-\underline{\pi }}\frac{\underline{\pi }}{\psi }
\left[ 1+\frac{p\left( \underline{\pi }\right) A_{q}(\underline{\pi })}{\phi }\right] =0 \label{pickySS} \\
\text{ }p(\underline{\pi })A_{q}(\underline{\pi }) &\geq& Q^{\ast }.  \notag
\end{eqnarray}
As $G\left( \underline{\pi },0,p\right) $ is linear in $p$, and strictly decreasing in $p$ since 
\begin{equation*}
\frac{A_{u}(\underline{\pi })}{\phi +(1-\phi )r} -\frac{1}{\phi } -\frac{1-\psi }{1-\underline{\pi }}\frac{\underline{\pi }}{\psi }\frac{A_{q}(\underline{\pi })}{\phi } < 0.
\end{equation*}
It follows that:
\begin{proposition}
	A two sector equilibrium with workers rejecting low tech jobs exists if and only if the following conditions hold:
	
	\begin{enumerate}
		\item $G\left( \underline{\pi },0,0\right) >0$
		
		\item $G\left( \underline{\pi },0,1\right) <0$
		
		\item $p\left( \underline{\pi }\right) A_{q}\left( \underline{\pi }\right) \geq Q^{\ast }$ for the solution to (\ref{pickySS}).
	\end{enumerate}
	
	There can be at most one such equilibrium.
\end{proposition}

Again, the proportion of high tech firms is the equilibrating variable, and, again, the steady state condition is linear in this proportion. Assuming that $\psi \geq \underline{\pi }$ we have that

\begin{equation}
G\left( \underline{\pi },0,0\right) =1+\frac{1}{\phi }-\frac{1-\psi }{1-\underline{\pi }}\frac{\underline{\pi }}{\psi }\geq \frac{1}{\phi },
\label{pos}
\end{equation}%
so a sufficient condition for $G\left( \underline{\pi },0,0\right) >0$ is that $\psi \geq \underline{\pi }.$ It is also the case that for any $\psi $ and $\underline{\pi }$ there exists $\phi $ sufficiently small for $G\left(\underline{\pi },0,0\right) >0.$ A small enough $\phi $ also ensures that $G\left( \underline{\pi },0,1\right) <0.$

\subsubsection{Workers Mixing}\label{twosectors_mix}

Finally, we consider the possibility of an equilibrium in which both sectors are active and $\underline{\pi }<\pi<\overline{\pi },$ which requires indifference on behalf of the workers. For indifference at some $\pi \in \left( \underline{\pi },\overline{\pi }\right) $ it is necessary that
\begin{equation}
p\left( \pi \right) =\frac{Q^{\ast }}{A_{q}\left( \pi \right) }\in \left[ 0,1\right] .
\end{equation}
Moreover, provided that $\underline{\pi }\leq \pi \leq \overline{\pi }$ there is a unique $\alpha \left( \pi \right) \in \left[ 0,1\right] $ that makes firms indifferent across sectors, $\alpha \left( \cdot \right) $ is strictly increasing, continuous and satisfies $\alpha \left( \underline{\pi }\right) =0$ and $\alpha \left( \overline{\pi }\right) =1.$ Hence, we
seek a solution so $\pi ^{\ast }\in \left( \underline{\pi },\overline{\pi } \right) $ to
\begin{equation} \label{mixingSS}
G\left( \pi ^{\ast },\alpha \left( \pi ^{\ast }\right) ,\frac{Q^{\ast }}{A_{q}\left( \pi ^{\ast }\right) }\right) =0. 
\end{equation}
Equilibria in which both firms and workers randomize may co-exist with other types of equilibria. However, we next establish that there is always at least one steady state equilibrium of the model.

\begin{proposition}
	\label{exist}There is always at least one equilibrium.
\end{proposition}

The proof is in the appendix.

\section{Numerical Examples}

To illustrate various possibilities, we now consider a pair of numerical examples. We set $f_q(\theta)=2\theta$, $f_u(\theta)=2(1-\theta)$, which is not only consistent with the monotone likelihood rate property, but also guarantees interior hiring thresholds as the likelihood ratio is 0 at $\theta=0$ and approaches infinity as $\theta$ converges to one. We also assume that $w_h$, $y_h$, $\beta$, $\phi$, and $r$ are set so that $W_q=1$ and $W_u=-1$ in each example. Solving for the optimal hiring threshold in \eqref{standard} we obtain $s(\pi)=1-\pi$. Plugging this threshold into $F_q(\theta)=\theta^2$ and $F_u(\theta)=2\theta-\theta^2$ and using the definitions in \eqref{Adef} we obtain $A_q(\pi)=\pi(2-\pi)$ and $A_u(\pi)=\pi^2$, respectively. All our numerical examples use these signals and firm values.

The purpose of the examples is to illustrate the difference between the model with and without group characteristics. In this section, in which have yet to introduce the group characteristic, we show that we may have unique of multiple equilibria. However, the example in which the equilibrium is unique is \textit{the exact same parametrization as the numerical example with multiple groups}. With multiple groups, asymmetric discriminatory equilibria appear, demonstrating how discrimination in this model is driven by spillover effects between groups as opposed to pure coordination.  

\subsection{Example 1: Unique Equilibrium} \label{exunique}
\begin{figure}[H]
	\centering
	\includegraphics[width=.5\linewidth]{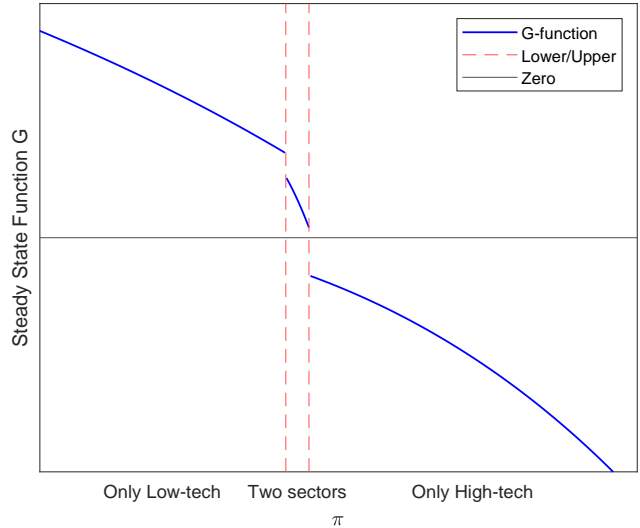}
	\caption{The bounds $\underline{\pi}=0.1647$ and $\overline{\pi}=0.1802$. Parameters used for this example are: $\beta=0.9, \phi=0.06, r=0.75, \psi=0.25, b=0.2$; and $W_q= 1$, $W_u=-1$, $y_l=0.5$, $w_l=0.495$. }
	\label{unique}
\end{figure}

Figure~\ref{unique} plots the relevant steady state equation as a function of $\pi$ for the three ranges in which $\pi$ is the variable that adjusts to support the steady state. For $\pi<\underline{\pi}$ we know that the only candidate solution is a low tech equilibrium with $\pi=\psi$, but we nevertheless plot $G(\pi,1,0)$ for completeness (the formula is in \eqref{lowSS}). For $\pi \in \left( \underline{\pi},\overline{\pi} \right)$, the equilibrium $\pi$ is pinned down by the steady state equation~\eqref{mixingSS}, and for $\pi \in \left(\overline{\pi},1 \right]$, the equilibrium $\pi$ is pinned down by the steady state equation~\eqref{highSS}. From Figure~\ref{unique} we see that there is neither a mixed, nor a high tech nor a low tech equilibrium.

\begin{figure}[H]
		\centering
		\includegraphics[width=.45\linewidth]{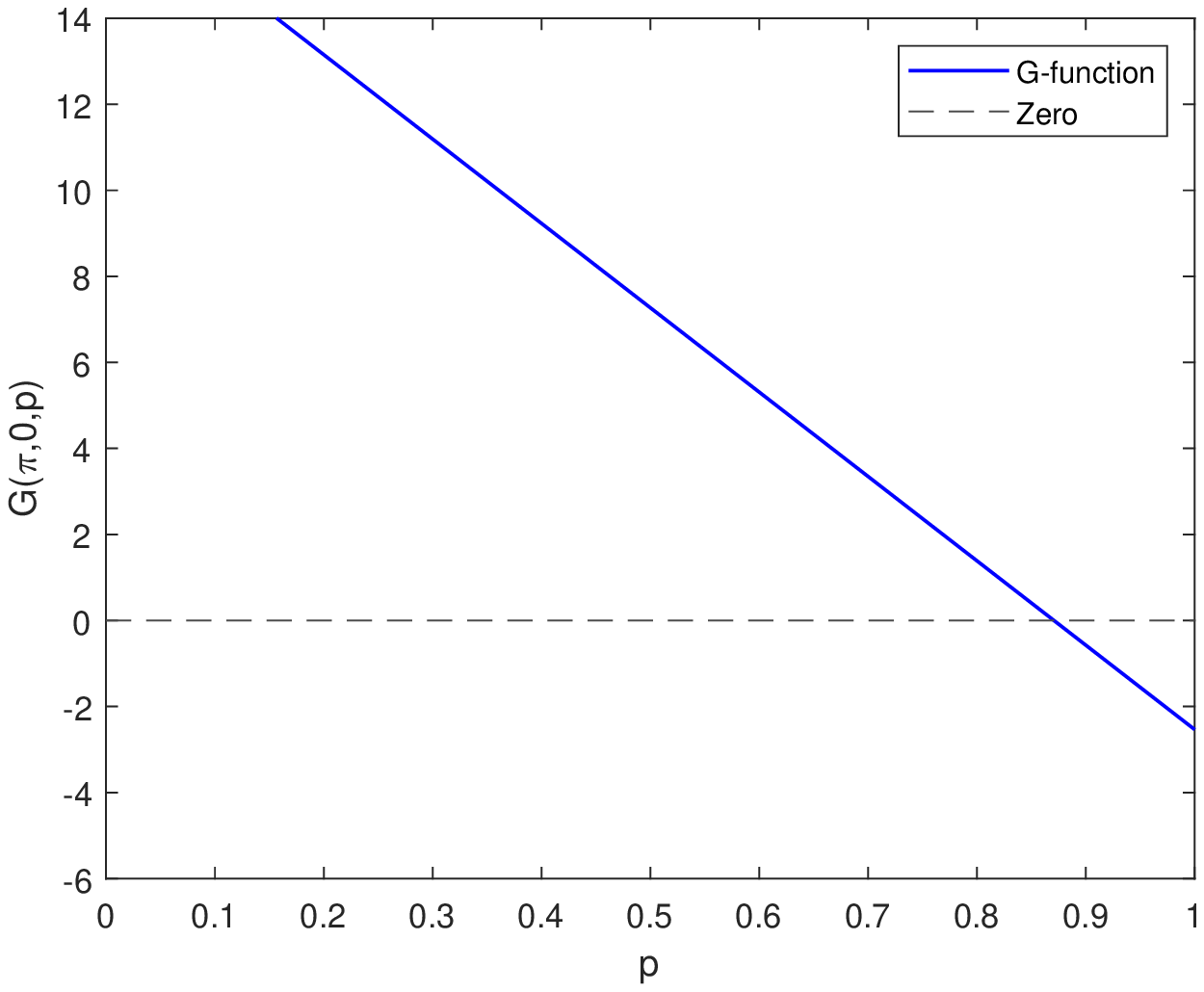}\hspace{.05\linewidth} \includegraphics[width=.45\linewidth]{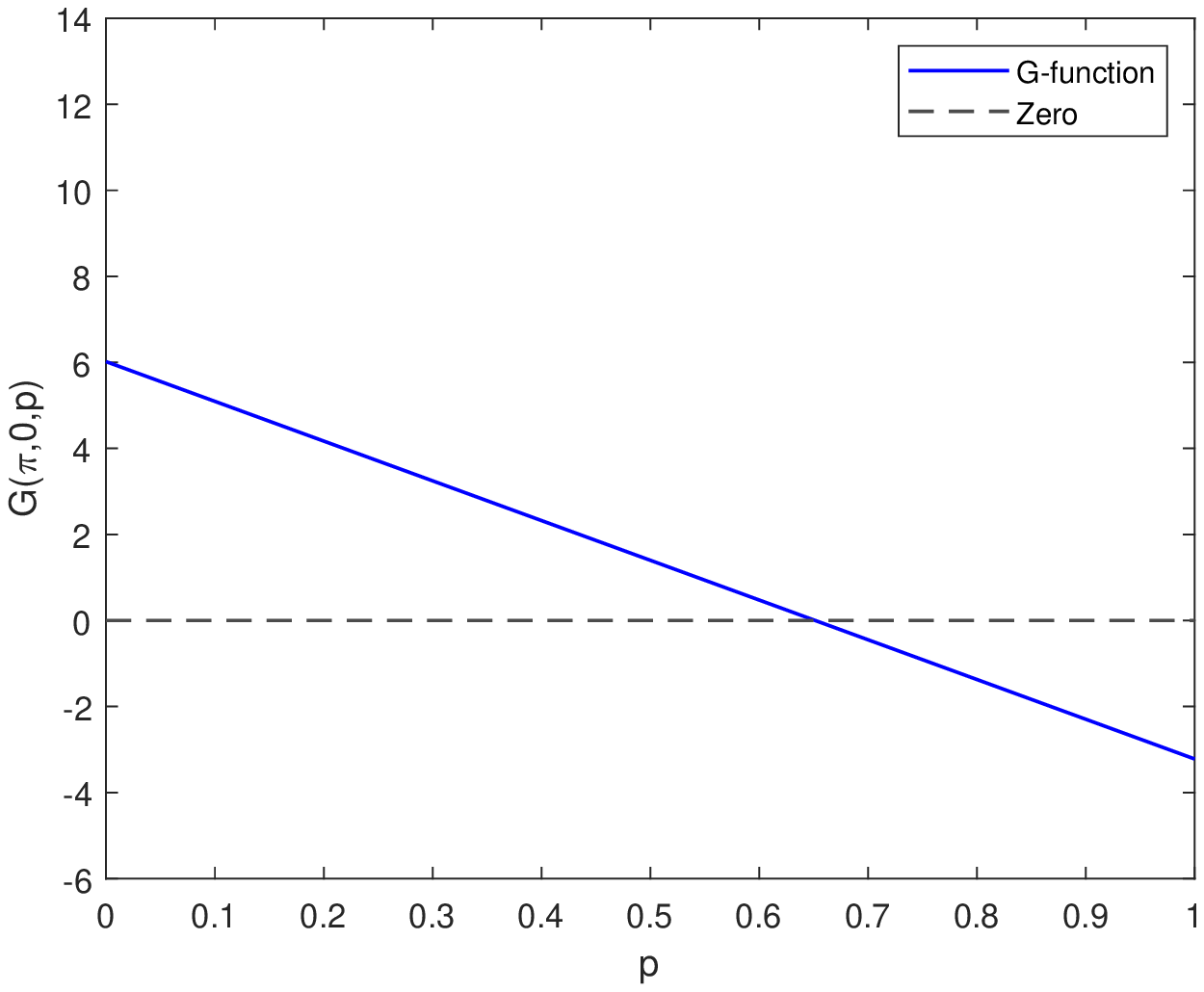}
		\caption{The left side shows the steady state value $p(\underline{\pi})$  and the right side shows the steady state value $p(\overline{\pi})$ The latter is not consistent with worker optimization.}
		\label{unique_Gp}
	\end{figure}

What cannot be seen from Figure~\ref{unique} is whether there is an equilibrium at $\pi=\underline{\pi}$ or $\pi=\overline{\pi}$. Solving for the unique $p(\underline{\pi})\in \left[ 0,1 \right]$ that is consistent with steady state in the left side of Figure~\ref{unique_Gp}, we find that under the parameter of the example $p(\underline{\pi})A_q(\underline{\pi})- Q^*= 0.8707\times 0.1647\times (2-0.1647) - 0.1830>0$, so rejecting low tech offers is consistent with worker optimality. However, at $\pi=\overline{\pi}$, the steady state value $p(\overline{\pi})\in\left[0,1 \right]$ shown in the right side of Figure~\ref{unique_Gp} is not consistent with workers' accepting low tech offers as $p(\overline{\pi})A_q(\overline{\pi}) - Q^*=0.6514\times  0.1802\times (2-0.1802) - 0.1830 > 0$. Hence there is a unique equilibrium in this example.

\subsection{Example 2: Multiple Equilibria}
\begin{figure}[H]
	\centering
	\includegraphics[width=.5\linewidth]{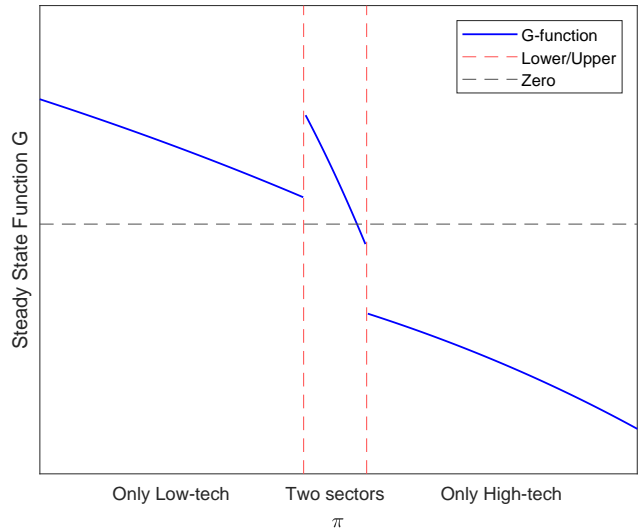}
	\caption{The bounds are $\underline{\pi}=0.2104$, $\overline{\pi}=0.2368$, and the equilibrium $\pi^*=0.2355$. Parameters used for this example are: $\beta=0.99, \phi=0.08, r=0.75, \psi=0.25, b=0.2$; and $W_q= 1$, $W_u=-1$, $y_l=0.5$, $w_l=0.495$.}
	\label{multiple_mixing}
\end{figure}

We now change the parameters so that the model admits multiple equilibria. See the appendix for details. Note that both sectors are active in every equilibrium of the example. The first equilibrium is a fully mixed equilibrium $\pi^*\in(\underline{\pi},\overline{\pi})$ shown in Figure~\ref{multiple_mixing}. Here both sectors are active, and qualified workers are indifferent between accepting and rejecting, so $\alpha(\pi^*)\in(0,1)$ is set to create indifference firms that enter.

\begin{figure}[H]
		\centering
		\includegraphics[width=.45\linewidth]{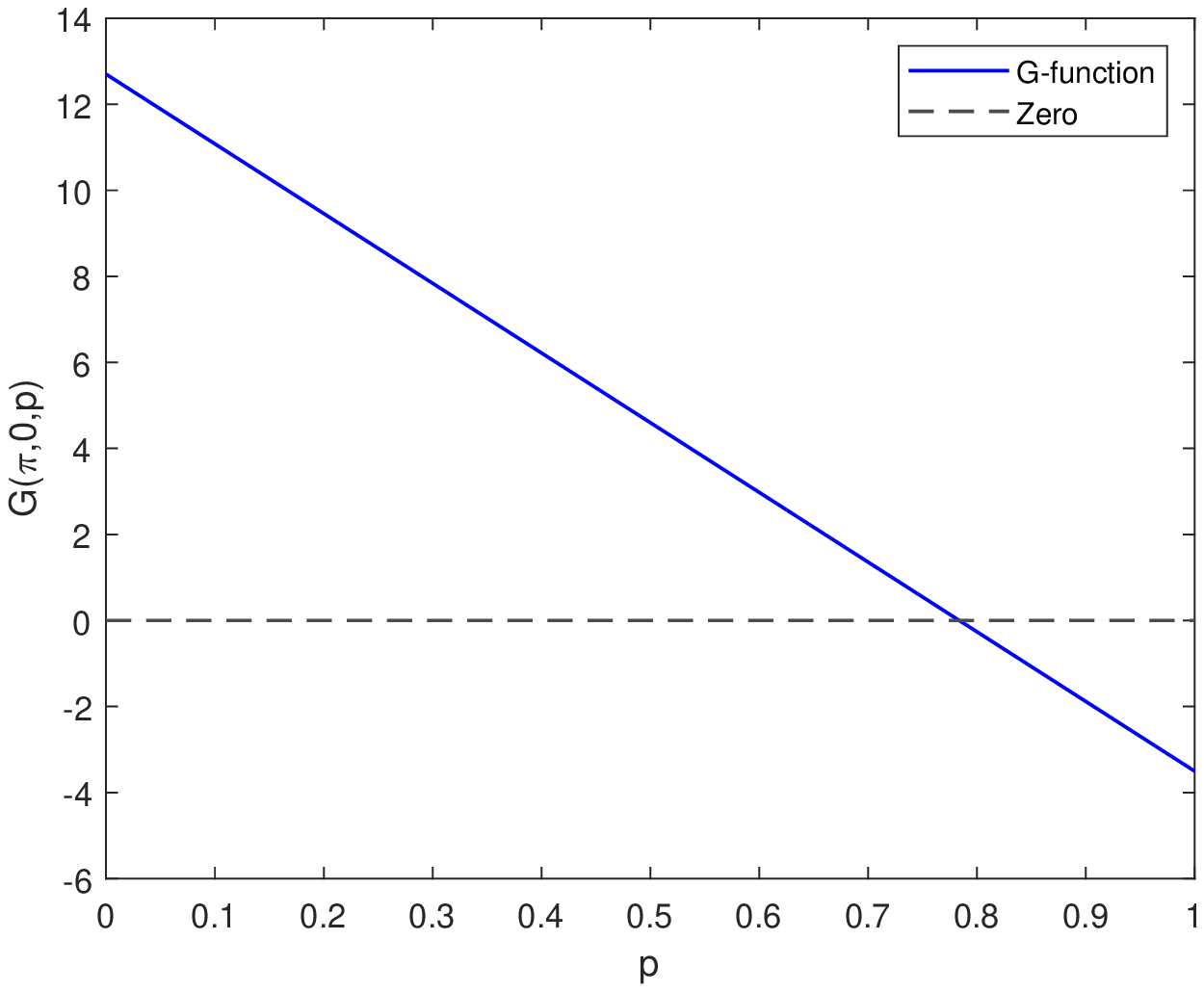}\hspace{.05\linewidth}
		\includegraphics[width=.45\linewidth]{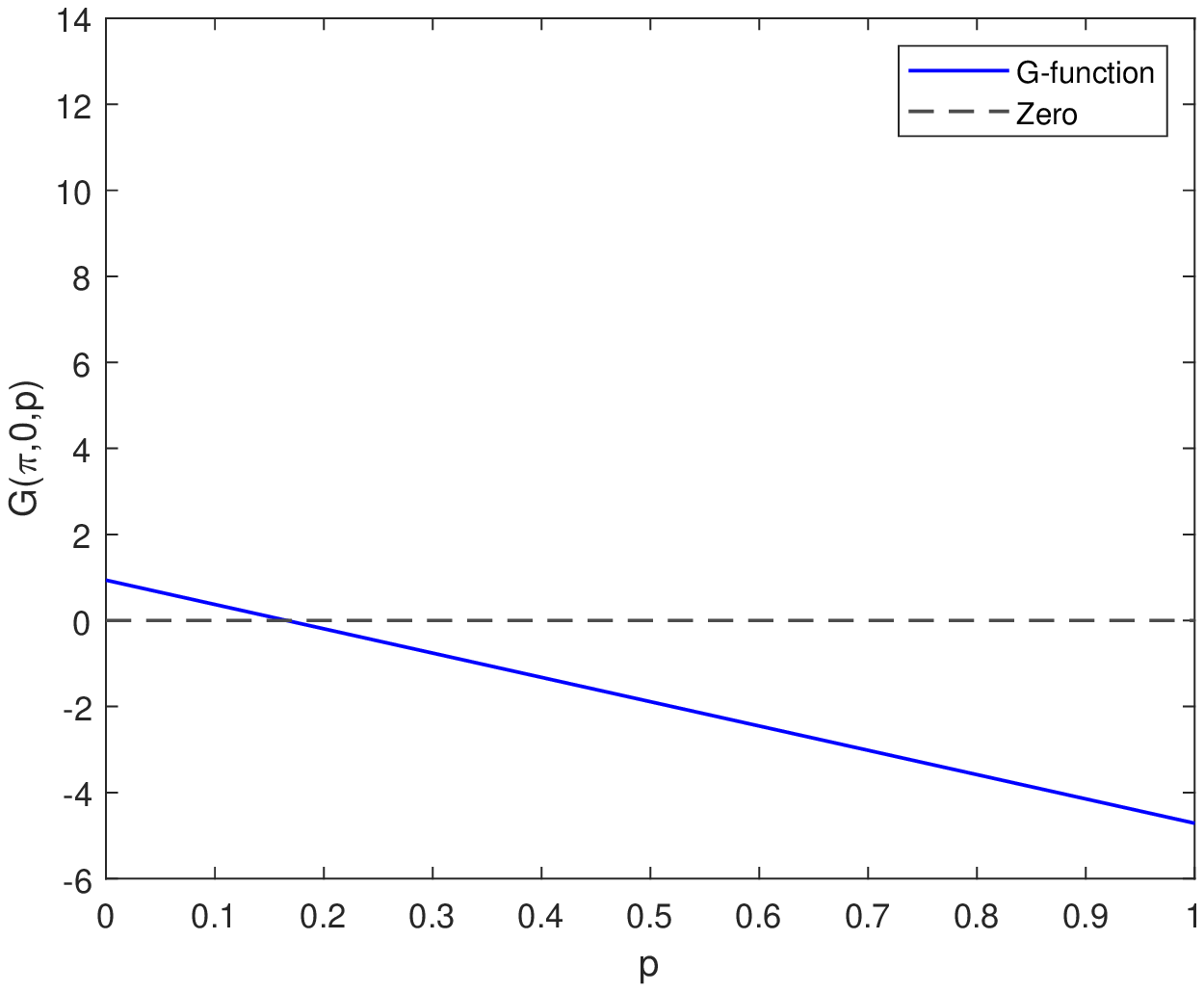}
		\caption{The left side shows the steady state equilibrium $p(\underline{\pi})$ at $\underline{\pi}$, and the right side shows the steady state equilibrium $p(\overline{\pi})$ at $\overline{\pi}$. }
		\label{multiple_Gp}
	\end{figure}

There are also two equilibria in which both sectors are active and workers are playing pure strategies at $\underline{\pi}$ and $\overline{\pi}$, respectively. At $\underline{\pi}$ low tech jobs must be rejected by workers to keep firms indifferent, so $\alpha(\underline{\pi})=0$. The equilibrium $p(\underline{\pi})$ characterized by \eqref{pickySS} is $p(\underline{\pi})=0.7838\in(0,1)$ shown in the left side of Figure (\ref{multiple_Gp}). And $p(\underline{\pi})A_q(\underline{\pi}) - Q^*= p(\underline{\pi})\underline{\pi}(2-\underline{\pi})-Q^* >0$ is consistent with workers' acceptance rule $\alpha(\underline{\pi})=0$.

At $\overline{\pi}$ low tech jobs must be accepted. The equilibrium $p(\overline{\pi})$ characterized by \eqref{notpickySS} is $p(\overline{\pi})=0.1658\in(0,1)$ shown in the right side of Figure~\eqref{multiple_Gp}. And $p(\overline{\pi}) A_q(\overline{\pi}) - Q^*= p(\overline{\pi}) \overline{\pi}(2-\overline{\pi})-Q^* < 0$ is consistent with workers' acceptance rule $\alpha(\overline{\pi})=1$.

\section{The Model with Observable Group Characteristics}

We now add a \textit{payoff irrelevant} group characteristic to the model.
Each worker belongs to group $j\in \{f,m \}$ and we denote by $\lambda^j$
the fraction of workers in the population that belongs to group $j$. For notational brevity, we
let $\boldsymbol{\pi }=\left( \pi ^{m},\pi ^{f}\right) $ denote the
endogenous stationary proportions of qualified workers from each group.

\subsection{Firm Optimization}

For the same reasons as in the baseline model,
the values of employing qualified and unqualified workers are independent of
$\boldsymbol{\pi }=\left( \pi ^{m},\pi ^{f}\right) $ and given by $W_{u}$
and $W_{q}$ defined in (\ref{constantfirmvalues}). Hence, there is no change
in the optimal firm hiring decision. Given group-specific prior $\pi ^{j}$
the and signal $\theta $ the firm is better off hiring the worker from group 
$j\in \{f,m\}$ in expectation if and only if 
\begin{equation}
P\left( \theta ,\pi ^{j}\right) W_{q}+\left( 1-P\left( \theta ,\pi
^{j}\right) \right) W_{u}\geq 0.
\end{equation}%
Hence, the optimal hiring rule is derived just like in the symmetric
model, so for any $j=m,f$ and any $\pi ^{j}\in \left[ 0,1\right] $ the
optimal hiring threshold is characterized as in Lemma \ref{cutoff}. We write $s\left( \pi ^{m}\right) $ and $s\left( \pi ^{f}\right) $ for
the thresholds as they can be determined independently. We also keep the
shorthand notation $A_{q}\left( \pi ^{j}\right) =1-F_{q}\left( s\left( \pi
^{j}\right) \right) $ for the probability that a qualified worker will get
an offer when matched with a high tech firm and $A_{u}\left( \pi ^{j}\right)
=1-F_{u}\left( s\left( \pi ^{j}\right) \right) $ for the corresponding probability for an unqualified worker. As in the symmetric model these are uniquely determined and
\begin{equation*}
\pi ^{j}A_{q}(\pi ^{j})W_{q}+(1-\pi ^{j})A_{u}(\pi ^{j})W_{u}
\end{equation*}
is strictly increasing in $\pi ^{j}$ for exactly the same reasons as the baseline model. 

The free entry conditions are the obvious extensions of the ones for the
symmetric model, namely 
\begin{eqnarray}
K &=&\beta p_{f}\sum_{j=f,m}\lambda ^{j}\left[ \pi ^{j}A_{q}(\pi
^{j})W_{q}+(1-\pi ^{j})A_{u}(\pi ^{j})W_{u}\right]  \\
K &=&\beta p_{f}\sum_{j=f,m}\lambda ^{j}\left[ \pi ^{j}\alpha ^{j}\left( 
\boldsymbol{\pi }\right) +(1-\pi ^{j})\right] W_{l},  \notag
\end{eqnarray}%
where $p_{f}$ is the probability that a firm matches with a worker, which,
since we use simple urn ball matching, is equated with the ratio of
unemployment over vacancies, or the inverse market tightness rate. Notice
that $\alpha ^{j}\left( \boldsymbol{\pi }\right) $, the randomization
probability for a worker in group $j$ is written as a function of the
proportion of qualified unemployed for\textit{\ both groups}. This is because
of feedback effects between groups that are explained below.

The argument is identical to the argument for the symmetric model. Also, we
note that a necessary condition for existence of an equilibrium with both sectors active is
\begin{eqnarray}
\sum_{j=f,m}\lambda ^{j}\left[ \pi ^{j}\alpha ^{j}\left( \boldsymbol{\pi}\right) +(1-\pi ^{j})\right] W_{l} =\sum_{j=f,m}\lambda ^{j}
\left[ \pi^{j}A_{q}(\pi ^{j})W_{q}+(1-\pi ^{j})A_{u}(\pi ^{j})W_{u}\right], \label{freeentry2}
\end{eqnarray}%
generalizing the condition in \eqref{freeentry} from the baseline model. This condition illustrates that if starting from a mixed strategy
equilibria, an increase in $\pi ^{j}$ no longer needs to be accompanied by
an increased probability of accepting low tech jobs \textit{for group ~}$j$
to keep firm indifference, as the other group can adjust their behavior
instead. Indeed, the worker optimization problem will create incentives for
such cross groups effects to be present.

\subsection{Worker Optimization}

As in the symmetric model, we only need to consider the qualified workers.
Denote by $V_{0}^{j}\left( \boldsymbol{\pi }\right) $, $V_{h}^{j}\left( \boldsymbol{\pi }\right) $ and $V_{l}^{j}\left( \boldsymbol{\pi }\right) $ the value for a qualified worker from being unemployed, being employed in the high tech sector, and being employed in the low tech sector, respectively.
Also, let $p\left( \boldsymbol{\pi }\right) $ be the probability that the worker meets a high tech firm, so that $1-p\left( \boldsymbol{\pi }\right) $ is the probability of meeting a low tech firm. Following the same steps (\eg  from the obvious extensions of (\ref{uvalue}) and (\ref{handlvalues}) as in the symmetric model we have that%
\begin{eqnarray}
V_{h}^{j}\left( \boldsymbol{\pi }\right)  &=&\frac{w_{h}+\phi \beta
	V_{0}^{j}\left( \boldsymbol{\pi }\right) }{1-\left( 1-\phi \right) \beta } \\
V_{l}^{j}\left( \boldsymbol{\pi }\right)  &=&\frac{w_{l}+\phi \beta
	V_{0}^{j}\left( \boldsymbol{\pi }\right) }{1-\left( 1-\phi \right) \beta }, 
\notag
\end{eqnarray}%
and we can characterize the worker optimality condition just like in the
symmetric model. Recall that $V^{\ast }$ is defined in (\ref{Vstar1}) as the
value of unemployment that makes the worker indifferent between accepting
and rejecting. For the same reasons as in the symmetric model we have that 
\begin{equation}
\alpha ^{j}\left( \boldsymbol{\pi }\right) =\left\{ 
\begin{array}{cc}
1 & \text{if }V_{0}^{j}\left( \boldsymbol{\pi }\right) <V^{\ast } \\ 
\left[ 0,1\right]  & \text{if }V_{0}^{j}\left( \boldsymbol{\pi }\right)
=V^{\ast } \\ 
0 & \text{if }V_{0}^{j}\left( \boldsymbol{\pi }\right) >V^{\ast }%
\end{array}
\right..
\end{equation}%
By substituting into the analog of (\ref{uvalue}) it follows that if the worker is indifferent, then it must be that
\begin{eqnarray}
V^{\ast }=p\left( \boldsymbol{\pi }\right) A_{q}\left( \pi ^{j}\right) \frac{ w_{h}+\phi \beta V^{\ast }}{1-\left( 1-\phi \right) \beta }
+\left(1-A_{q}\left( \pi ^{j}\right) p\left( \boldsymbol{\pi }\right) \right) \left( b+\beta V^{\ast }\right).
\end{eqnarray}%
Hence, the characterization of the optimal worker behavior is \textit{almost}
identical to the symmetric model. Using the exact same steps as in the baseline model, we have that:

\begin{lemma}
	Suppose that $0 \leq w_l-b \leq \beta (1-\phi) (w_h - b)$ and let $Q^{\ast }\in \left[ 0,1\right] $
	be defined in Lemma \ref{workeroptimal}. Then, the optimal worker choice correspondence for group $j$ is, 
	\begin{equation}
	\alpha ^{j}\left( \boldsymbol{\pi }\right) =\left\{ 
	\begin{array}{cc}
	1 & \text{if }p\left( \boldsymbol{\pi }\right) A_{q}\left( \pi ^{j}\right) <Q^{\ast } \\ 
	\left[ 0,1\right]  & \text{if }p\left( \boldsymbol{\pi }\right) A_{q}\left(\pi ^{j}\right) =Q^{\ast } \\ 
	0 & \text{if }p\left( \boldsymbol{\pi }\right) A_{q}\left( \pi ^{j}\right) >Q^{\ast }
	\end{array}%
	\right. .  \label{acceptgroup}
	\end{equation}
\end{lemma}

At this point, it may appear that the two group model collapses to the baseline model, but this isn't so because of a subtle difference. We assume that workers from the two groups are randomly matched with firms. Hence, both groups are faced with the same probability of matching a high tech firm, implying that firm entry cannot adjust to keep both groups willing to randomize unless $\pi ^{m}=\pi ^{f}.$

\subsection{The Fragility of Mixing with Multiple Groups}

In the symmetric model, should $\pi$ change slightly from a fully mixed equilibrium, an associated small change in the proportion of high tech firms can restore indifference for the workers. In contrast, when there are multiple groups and $\pi^j$ is perturbed from a symmetric mixed strategy equilibrium, there is simply no way for the proportion of high tech firms to adjust so as to make both groups indifferent. This is immediate from the acceptance rules in (\ref{acceptgroup}). It follows that the response must involve at least one group to reject low tech offers for sure or one group to accept low tech offers for sure. Hence, equilibria in which both groups are mixing is a knife edge possibility, whereas mixing in the symmetric model may be a robust possibility.

\subsection{Steady State Conditions}

Refer to the symmetric model and note that because the steady state
conditions in terms of $p$, $\alpha ^{j}$ and $\pi ^{j}$ are just like the
symmetric model, we now how a pair of steady state conditions%
\begin{eqnarray}
G( \pi ^{f},\alpha ^{f}\left( \boldsymbol{\pi }\right) ,p\left( 
\boldsymbol{\pi }\right) )  &=&0  \label{pairSS} \\
G\left( \pi ^{m},\alpha ^{m}\left( \boldsymbol{\pi }\right) ,p\left( 
\boldsymbol{\pi }\right) \right)  &=&0,  \notag
\end{eqnarray}%
where  $G$ is defined in (\ref{Gdef}). Derivations are identical to the
baseline model. Hence:

\begin{definition}
	A steady state equilibrium is an object $\left(\boldsymbol{\pi}, \alpha^f(\boldsymbol{\pi}), \alpha^m(\boldsymbol{\pi}),  p(\boldsymbol{\pi}) \right)$ such that (\ref{pairSS}) holds, $\left(\alpha ^{f}\left( \boldsymbol{\pi}\right),\alpha ^{m}\left(\boldsymbol{\pi }\right) \right)$ satisfy the worker optimality condition (\ref{acceptgroup}) and $p\left( \boldsymbol{\pi }\right) $ is consistent with optimal entry.
\end{definition}

It is immediate from (\ref{pairSS}) that if $\left( \pi ,\alpha \left( \pi\right) ,p\left( \pi \right) \right) $ is an equilibrium in the baseline model, then $\left( \pi ,\pi \right) $ satisfies (\ref{pairSS}). Moreover, the optimality condition (\ref{acceptgroup}) and firm entry conditions reduce to the ones in the baseline model. Hence, any equilibrium in the baseline model corresponds to a non-discriminatory equilibrium in the model with observable group characteristics $\left(\boldsymbol{\pi },\alpha ^{f}\left( \boldsymbol{\pi }\right),\alpha ^{m}\left( \boldsymbol{\pi }\right) ,p\left( \boldsymbol{\pi }\right)\right) =\left( \left( \pi ,\pi \right), \alpha \left( \pi \right),\alpha \left( \pi \right), p\left( \pi \right) \right)$.

\section{Equilibria with Discrimination}

We now argue that even in the case with a unique equilibrium in the symmetric model, there may be asymmetric equilibria in the model with
observable group characteristics. 

Assume that $\underline{\pi }<\Pi ^{\ast }<\overline{\pi }$ is such that $A_{q}\left( \Pi ^{\ast }\right) =Q^{\ast }.$ Since $Q^{\ast }$ can be set in any way we want without affecting the incentives for the firms by simultaneously changing the wage and the productivity this is always
possible. Also assume that $\psi >\overline{\pi },$ which assures that $G\left( \overline{\pi },1,0\right) >0.$ Moreover, let $\phi $ be small enough so that any $\pi ^{\ast }<\Pi ^{\ast }$ for any $\pi ^{\ast }$ such that $\widetilde{G}\left( \pi ^{\ast }\right) =0$ and so that $G\left( \overline{\pi },1,1\right) <0$ and $p\left( \overline{\pi }\right) <\frac{Q^{\ast }}{A_{q}\left( \overline{\pi }\right) }$ for the unique solution to $G\left( \overline{\pi },1,p\left( \overline{\pi }\right) \right) =0.$ This is possible as $\pi ^{\ast }\left( \phi \right) \rightarrow 0$ as $\phi
\rightarrow 0$ for any sequence of solutions to $\widetilde{G}\left( \pi^{\ast }\left( \phi \right) \right) =0,$ $G\left( \overline{\pi },1,1\right) \rightarrow -\infty $ as $\phi \rightarrow 0$ and $p\left( \overline{\pi } \left( \phi \right) \right) \rightarrow 0$ as $\phi \rightarrow 0$ for any sequence of solutions to $G\left( \overline{\pi },1,p\left( \overline{\pi };\phi \right) \right) =0.$ Together these assumptions rule out any kind of symmetric equilibrium except for a fully randomized equilibrium.

Moreover, we have that 
\begin{equation}
\widetilde{G}\left( \Pi ^{\ast }\right) =G\left( \Pi ^{\ast },\alpha \left(\Pi ^{\ast }\right) ,1\right) <0<G\left( \overline{\pi },1,\frac{Q^{\ast }}{A_{q}\left( \overline{\pi }\right) }\right)   \label{conda}
\end{equation}
where the first inequality is because there is no high tech equilibrium and the second because there is no two sector equilibrium at $\overline{\pi }.$ By continuity, there exists some fully mixed symmetric equilibrium $\pi^{\ast }\in \left( \Pi ^{\ast },\overline{\pi }\right) .$ In general, it may or may not be unique, but the case with uniqueness (which is possible) is the more interesting case. 

\begin{proposition}
	\label{discprop}Suppose that (\ref{conda}) is satisfied. Then there exists a fully mixed symmetric equilibrium $\left( \pi ^{\ast },\alpha \left( \pi^{\ast}\right) ,\frac{Q^{\ast }}{A_{q}\left( \pi ^{\ast }\right) }\right).$ Moreover, for any  such symmetric equilibrium there exists an interval $\left( \underline{p},\overline{p}\right) $ containing $\frac{Q^{\ast }}{A_{q}\left( \pi ^{\ast }\right) }$ and $\pi ^{f}\left( p\right) <\pi ^{\ast}<\pi ^{m}\left( p\right) $ such that each $p\in \left( \underline{p},\overline{p}\right) ,$ $\left( \pi ^{f}\left( p\right) ,\pi ^{m}\left(p\right) \right) $ corresponds to an asymmetric equilibrium in which $\alpha^{f}\left( p\right) =1$ and $a^{m}\left( p\right) =0$ for some population proportions $\left( \lambda ^{f}\left( p\right) ,\lambda ^{m}\left( p\right) \right) $ with $\lambda ^{f}\left( p\right) +\lambda ^{m}\left( p\right) =1.$ Moreover, if $\phi $ is small enough $\lambda ^{m}\left( p\right) $ is
	strictly increasing in $p$ implying that there is a generic set of
	population fractions such that an asymmetric equilibrium exists.
\end{proposition}

The idea is straightforward, but some of the details of the proof in the appendix are somewhat tedious. The first step simply notes that if the proportion of high tech firms stays the same and women accept low tech jobs for sure and men reject them for sure, then the steady state proportions of qualified men and females diverge. Men are now more likely to be qualified and women are less likely to be qualified than in the randomized equilibrium. All else equal the profitability of meeting men (women) increases (decreases) in the high tech sector and decreases (increases) in the low tech sector, so the population proportion that leaves the firms indifferent at the original high tech firm probability $Q^{\ast}/A_{q}\left( \pi ^{\ast }\right) $ are uniquely determined. However, $p$ can be perturbed around $Q^{\ast }/A_{q}\left( \pi ^{\ast }\right) $ while still having men with a strict incentive to reject low tech jobs and women having a strict incentive to accept. This can be used to show that there is a robust set of $\left( \lambda ^{f},\lambda ^{m}\right) $ for which an asymmetric equilibrium exists.

It is important to notice that nowhere is the proof of Proposition \ref{discprop} relying on multiplicity in the underlying one-group model. Instead, the result is driven by the fact that men and women compete \textit{in the same market}, which allows the two groups to specialize in equilibrium.\footnote{This is somewhat similar to \cite{Bardhi} where learning dynamics can create sizable inequality from small differences between groups} All else equal, if men get pickier, which increases the fraction of qualified unemployed men, the high tech sector gets more profitable. To restore equal profits, it is thus necessary for women to get less picky. One could, of course, object that instead the high tech sector should compete away the low tech jobs, but this would drive down the proportion of qualified workers in both groups to such an extent that only low tech firms would like to enter when $\phi $ is small.

\subsection{Numerical Example of Discrimination}
We consider the two group version of the model in Section~\ref{exunique}, assuming groups are of equal size. Recall that in the baseline model, this is a parametrization in which the equilibrium is unique. This equilibrium corresponds to a symmetric equilibrium with two groups, but as we show below, there are now also discriminatory equilibria, illustrating that the model creates potential incentives for specialization.

We know from from~\eqref{acceptgroup} that at most group can randomize. There may be equilibria in which no group randomizes, but we will consider the case in which $\pi^m>\pi^f$ and $\alpha^f(\boldsymbol{\pi})\in(0,1)$. Then, we have that $\alpha^m(\boldsymbol{\pi})= 0$ from~\eqref{acceptgroup}:
\begin{equation*}
\underbrace{p\left( \boldsymbol{\pi }\right) A_{q}\left(\pi ^{m}\right) > p\left( \boldsymbol{\pi }\right) A_{q}(\pi ^{f} ) = Q^{\ast } }_{\text{hiring probability of high tech firm}} \text{ and } \underbrace{\alpha^m = 0<\alpha^f\in (0,1)}_{\text{acceptance of low tech job}}.
\end{equation*}
This can be interpreted as a \textit{cross group effect} that comes from both groups searching in the same labor market. They therefore share  the same probability of matching with a high tech firm $p(\boldsymbol{\pi})$, which drives the spillovers across groups.
    	   
In the discriminatory equilibrium $p(\boldsymbol{\pi })A_q(\pi^m)> Q^*= p(\boldsymbol{\pi })A_q(\pi^f)$ to justify the worker acceptance rules. Moreover, the indifference  condition for two active sectors (\ref{freeentry2}) evaluated at $\alpha^m(\boldsymbol{\pi})= 0$ simplifies to 
\begin{eqnarray}
\lambda^m (1-\pi^{m}) W_{l}+\lambda^f \left[ \pi^f \alpha^f(\boldsymbol{\pi}) +(1-\pi^f) \right] W_{l}
=\sum_{j=f,m} \lambda^j \left[ \pi ^{j}A_{q}(\pi^{j})W_{q}+(1-\pi^{j})A_{u}(\pi ^{j})W_{u}\right]. \label{firmindifference2}
\end{eqnarray}
This condition pins down the female group's acceptance decision $\alpha^f(\boldsymbol{\pi})$ as an increasing function in both $\pi^f$ and $\pi^m$,
\begin{equation}\label{acceptfemale}
	\alpha^f(\boldsymbol{\pi})=\frac{\sum_{j=f,m} \lambda^j \left[ \pi ^{j}A_{q}(\pi ^{j})W_{q}+(1-\pi^{j})A_{u}(\pi ^{j})W_{u}-(1-\pi^j)W_l \right] }{\lambda^f \pi^f W_l}.
\end{equation}

The steady state conditions (\ref{pairSS}) evaluated at $\alpha^m(\boldsymbol{\pi})=0$ and $p(\boldsymbol{\pi})=\frac{Q^*}{A_q(\pi^f)}$ are then
\begin{eqnarray}
&&1+ \frac{Q^*}{A_q(\pi^f)} \frac{ A_{u}(\pi^m)}{\phi +(1-\phi )r}+\frac{1-\frac{Q^*}{A_q(\pi^f)} }{\phi} =\frac{1-\psi }{1-\pi ^m }\frac{\pi ^m }{\psi }
\left[ 1+ \frac{Q^*}{A_q(\pi^f)}  \frac{A_{q}(\pi^m )}{\phi } \right], \label{maleSS} \\
&&1+\frac{Q^*}{A_q(\pi^f)} \frac{A_{u}(\pi ^f )}{\phi +(1-\phi)r} +\frac{1- \frac{Q^*}{A_q(\pi^f)} }{\phi } =\frac{1-\psi }{1-\pi^f }\frac{\pi^f }{\psi } \left[ 1+ \frac{Q^*}{\phi} +\frac{ 1-\frac{Q^*}{A_q(\pi^f)}  }{\phi } \alpha^f(\boldsymbol{\pi})  \right], \label{femaleSS}
\end{eqnarray}
where $\alpha^f(\boldsymbol{\pi})$ is determined in~\eqref{acceptfemale}. These two equations pin down the equilibrium $\left\{\pi^m,\pi^f \right\}$. If the group $m$ workers' condition $ p(\boldsymbol{\pi})A_q(\pi^f)=Q^*<p(\boldsymbol{\pi}) A_q(\pi^m)$ hold, then the equilibria candidate is an equilibrium.	

\begin{figure}[H]
	\centering
	\includegraphics[width=.5\linewidth]{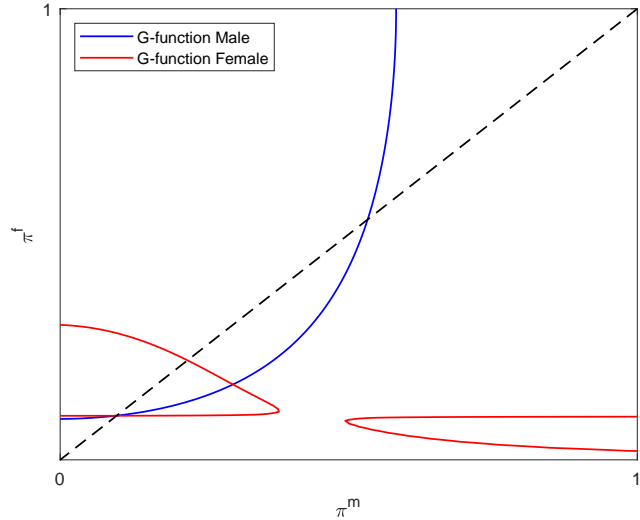}
	\caption{Discrimination equilibria with $\pi^m>\pi^f$. The red line captures the steady state function for women, the blue line captures the steady state function for men. Parameters for this example are the same as Section~\ref{exunique}: $\beta = 0.9$, $\phi = 0.06$, $r = 0.75$, $\psi = 0.25$, $b = 0.2$.}
	\label{discfig}
\end{figure}

Substituting the parametric assumptions into equations~\eqref{maleSS} and~\eqref{femaleSS} we plot the result in Figure~\ref{discfig}. The plot shows the discrimination equilibrium beliefs about the qualification of group $f$ and $m$. The steady state condition for male group~\eqref{maleSS} is captured by the blue line, and the steady state condition for female group~\eqref{femaleSS} is represented by the red line. The equilibrium beliefs about the qualification of the male group and female group $\pi^m>\pi^f$, confirming that we have constructed an equilibrium.

In the discriminatory equilibrium, observing a signal about workers' qualification, high tech firms set a higher hiring threshold for the workers from group $f$. Hence, high tech firms offer a job to the workers from group $f$ with a lower probability than workers from group $m$. As a result, group $f$ workers incentives to accept low tech jobs are reinforced. One may notice that this looks a lot like women are less self-confident than men, so our model could be viewed as an instrumental model of effects of confidence on the gender gap studied by \citet{kamas2018competing} and others.

\section{Discussion: Affirmative Action}

Most applied theory papers on affirmative action are aimed at pointing out potential unintended effects of the policy. \citet{coate1993will} argue
that the policy may need to be permanent because incentives to acquire human capital may be perversely affected. \citet{moro2003affirmative} point out that
if wages are endogenous, affirmative action may not even benefit the targeted group. \citet{chan2003does} show that banning affirmative action in
college admissions may backfire because universities create more randomness in the admission process. Finally, \cite{Fershtmanpavan} argue that requiring a larger proportion of minority candidates to be considered can backfire as employers may respond by increasing the pool of candidates.

In our model, when comparing steady states before and after the policy, none of these unintended consequences occur. Assuming that the policy mandates
that the proportion of men and women should be equalized in the high tech sector, the only possible steady state equilibrium is a symmetric equilibrium. Should $\pi ^{f}<\pi ^{m}$, the only way the firms could satisfy the quota is to be less demanding on female applicants. Hence, women would have weaker incentives to accept low tech jobs, which is inconsistent with $\pi ^{f}<\pi ^{m}$. While we have not worked out the adjustment dynamics, this suggests that gender quotas could be useful if statistical discrimination driven by occupational choice is important for the gender wage gap.

Our model and \citet{coate1993will} interpret affirmative action as a relative hiring quota, and the two models have a similar externality. In \citet{coate1993will}, the more your group invests the higher is the prior belief about the worker, while in our case, the pickier your group is the higher is the prior belief about the worker. So, why are the results so different? Besides our model being a full-fledged dynamic model, the key difference is that making it easier to get the more attractive jobs unambiguously creates the right incentives to reject low end jobs. In contrast, \citet{coate1993will} focuses on the possibility that making it easier to get the attractive jobs distorts incentives for pre-market investments in human capital.

\section{Conclusion}
We propose a search model of statistical discrimination. The interaction between occupational choice, search externalities, and a signal extraction problem makes it possible that identical groups specialize in equilibrium, resulting in cross-group inequality. Groups share the same labor market, creating spillover effects between groups, and discriminatory equilibria may exist also when the baseline model without group characteristic has a unique equilibrium. For the same reason, it is possible that the introduction of group characteristics destabilizes a symmetric equilibrium. Unlike the previous literature, affirmative action is an appropriate remedy to eliminate discriminatory equilibria in our model.

\pagebreak

\appendix

\section{Appendix A: Omitted Proofs}

\subsection{Proof of Lemma~\ref{cutoff}}\label{app_cutoff}

\begin{proof}
	Suppose that $s\left(\pi\right) $ solves $P\left( s\left(\pi \right) ,\pi\right) W_{q}+\left( 1-P\left( s\left( \pi \right) ,\pi \right) \right)W_{u}=0,$ which can be rearranged as (\ref{standard}). By the monotone likelihood property it follows that $P\left(\theta, \pi \right) W_{q} + \left( 1-P\left(\theta, \pi \right) \right) W_{u}>0$ for $\theta >s\left( \pi \right) $ and $P\left( \theta ,\pi \right) W_{q}+\left(1-P\left( \theta ,\pi \right) \right) W_{u} <0$ for $\theta <s\left( \pi\right) $, so the firm has a strict incentive to hire a worker with $\theta>s\left( \pi \right) $ and a strict incentive not to hire a worker with $\theta <s\left( \pi \right).$ The corner cases are immediate.
\end{proof}

\subsection{Proof of Lemma~\ref{profit}}\label{app_profit}

\begin{proof}
Applying the envelope theorem to
	$\pi A_{q}(\pi )W_{q}+(1-\pi )A_{u}(\pi )W_{u} = \max\limits_{\theta }\pi \left[1-F_{q}(\theta )\right] W_{q}+(1-\pi )\left[ 1-F_{u}(\theta )\right] W_{u} ,$ we have
	\begin{eqnarray}
	&&\frac{d }{d\pi}\left[ \pi A_q(\pi) W_q+ (1-\pi) A_u(\pi) W_u\right] \notag \\
	&=& \left[1-F_q(\theta(\pi)) \right] W_q -
	\left[1-F_u(\theta(\pi))\right] W_u>0, \notag
	\end{eqnarray}
	where the strict inequality follows from $1-F_{q}(\theta )>1-F_{u}(\theta ),$ which is an implication of $\frac{f_{q}}{f_{u}}$ being strictly increasing in $\theta $, and $W_{q}>W_{u}$.
\end{proof}

\subsection{Proof of Lemma~\ref{firmindifference}}\label{app_firmindifference}
\begin{proof}
    \noindent \textbf{To show (1).} For $\pi< \underline{\pi}$,
	\begin{eqnarray}
	&& -K+ \beta p_f \left[ \pi A_q(\pi) W_q + (1-\pi) A_u(\pi) W_u  \right] \notag \\
	&<& -K+ \beta p_f \left[ \underline{\pi} A_q(\underline{\pi}) W_q + (1-\underline{\pi}) A_u(\underline{\pi}) W_u  \right] \notag \\
	&=& -K+ \beta p_f \left( 1-\underline{\pi} \right) W_l \notag \\
	&<& -K+ \beta p_f \left( 1-\pi \right) W_l\notag \\
	&\leq& -K+ \beta p_f \left[ \pi \alpha + \left( 1-\pi \right)\right] W_l\notag 	
	\end{eqnarray}
	where the first inequality follows from $\pi A_q(\pi) W_q + (1-\pi) A_u(\pi) W_u$ is strictly increasing (in Lemma~\ref{profit}), the equality follows from $(1-\underline{\pi})W_l=\underline{\pi} A_q(\underline{\pi})W_q + (1-\underline{\pi}) A_u(\underline{\pi})W_u$, and the last inequality hold for any $\alpha\in[0,1]$. Hence, the low tech sector dominates high-tech sector for any $\alpha \in [0,1]$ when $\pi\in [0,\underline{\pi})$. 
	
	\medskip
	\noindent \textbf{To show (5).} For $\pi> \overline{\pi}$,
	\begin{eqnarray}
	&& -K+ \beta p_f \left[ \pi A_q(\pi)W_q+(1-\pi) A_u(\pi) W_u  \right] \notag \\
	&>& -K +\beta p_f\left[ \overline{\pi}  A_q(\overline{\pi}) W_q+ (1-\overline{\pi}) A_u(\overline{\pi}) W_u \right] \notag \\ 
	&=& -K+ \beta p_f  W_l \notag \\
	&\geq& -K+ \beta p_f \left[\pi\alpha+(1-\pi)\right] W_l. \notag 
	\end{eqnarray}
	where the first inequality follows from $\pi A_q(\pi) W_q + (1-\pi) A_u(\pi) W_u$ is strictly increasing (in Lemma~\ref{profit}), the equality follows from $W_l = \overline{\pi} A_q(\overline{\pi}) W_q + (1-\overline{\pi}) A_u(\overline{\pi}) W_u $, the second inequality holds for any $\alpha\in[0,1]$. Hence, entering the high-tech sector dominates the low tech sector for any $\alpha \in [0,1]$ when $\pi \in (\overline{\pi},1]$. 
	
	\medskip
	\noindent \textbf{To show (2) (3) (4).} The workers' acceptance rule $\alpha$ matters only when two sectors are active. When both sectors are active, the indifference condition holds
	\begin{equation*}
	\left[ \pi \alpha \left( \pi \right) +(1-\pi )\right] W_{l}=\pi A_{q}(\pi
	)W_{q}+(1-\pi )A_{u}(\pi )W_{u},
	\end{equation*}
	from which we have a unique $\alpha(\pi)=\frac{\pi A_{q}(\pi)W_{q}+(1-\pi )A_{u}(\pi )W_{u}-(1-\pi ) W_{l} }{\pi W_l}$. 
	
	From Lemma~\ref{profit}, $\pi  A_q(\pi) W_q+ (1-\pi) A_u(\pi) W_u$ is strictly increasing. It is thus immediate that $\alpha(\pi)$ must be strictly increasing to keep firms indifferent between the two sectors.
	
\end{proof}

\subsection{Proof of Lemma~\ref{workeroptimal}}

\begin{proof}
For the first part, equation (\ref{Q}) is simply a linear function of $Q\left( \pi \right) $. It has solution 
\begin{equation*}
Q^{\ast }=\frac{\frac{w_{l}-(1-(1-\phi )\beta )b}{(1-\phi )\beta }-w_{l}}{w_{h}-w_{l}}\in \left[0,1\right],
\end{equation*}
provided that $0\leq w_{l}-b\leq \beta (1-\phi )(w_{h}-b)$. For the second part, suppose that $\alpha \left( \pi \right) =1$ is optimal. Then, by (\ref{acceptreject2}) $V_{0}(\pi )\leq V^{\ast }$ and, directly from (\ref{uvalue}) $V_{l}(\pi )\geq b+\beta V_{0}(\pi )$. Equation (\ref{uvalue}) becomes 
\begin{eqnarray}
V_{0}(\pi ) &=&p\left( \pi \right) \left[ A_{q}\left( \pi \right) \frac{w_{h}+\phi \beta V_{0}(\pi )}{1-\left( 1-\phi \right) \beta }+\left(1-A_{q}\left( \pi \right) \right) \left( b+\beta V_{0}(\pi )\right) \right]
 +\left( 1-p\left( \pi \right) \right) V_{l}\left( \pi \right)   \notag \\
&\geq& p\left( \pi \right) \left[ A_{q}\left( \pi \right) \frac{w_{h}+\phi \beta V_{0}(\pi )}{1-\left( 1-\phi \right) \beta } +\left( 1-A_{q}\left( \pi \right) \right) \left( b+\beta V_{0}(\pi )\right) \right]
+\left( 1-p\left(\pi \right) \right) \left( b+\beta V_{0}(\pi )\right)  \notag \\
&=& p\left( \pi \right) A_{q}\left( \pi \right) \frac{w_{h}+\phi \beta V_{0}(\pi )}{1-\left( 1-\phi \right) \beta }+\left( 1-A_{q}\left( \pi \right) p\left( \pi \right) \right) \left( b+\beta V_{0}(\pi )\right)   \notag \\
&=& p\left( \pi \right) A_{q}\left( \pi \right) \left[ \frac{w_{h}+\phi \beta V_{0}(\pi )}{1-\left( 1-\phi \right) \beta }-\left( b+\beta V_{0}(\pi)\right) \right] +\left( b+\beta V_{0}(\pi )\right) 
\end{eqnarray}
or 
\begin{equation*}
p\left( \pi \right) A_{q}\left( \pi \right) \leq \frac{V_{0}(\pi )-\left(b+\beta V_{0}(\pi )\right) }{\frac{w_{h}+\phi \beta V_{0}(\pi )}{1-\left(1-\phi \right) \beta }-\left( b+\beta V_{0}(\pi )\right) },
\end{equation*}
where the RHS is increasing in $V_{0}(\pi )$ and $V_{0}(\pi )\leq V^{\ast }$. Hence, $p\left( \pi \right) A_{q}\left( \pi \right) \leq Q^{\ast }$. A symmetric argument shows that $\alpha \left( \pi \right) =0$ is optimal if and only if $p\left( \pi \right) A_{q}\left( \pi \right) \geq Q^{\ast }$ and the derivation of (\ref{Q}) shows that the worker is indifferent if and only if $p\left( \pi \right) A_{q}\left( \pi \right) =Q^{\ast }$. 
\end{proof}

\subsection{Proof of Proposition~\ref{loweq}}
\begin{proof}
Suppose an equilibrium with $p\left( \pi \right) =0$ exists, which implies that $\alpha(\pi)=1$ from (\ref{acceptreject3}). For this to
be consistent with the steady state conditions%
\begin{equation}
G\left( \pi ,1,0\right) =1+\frac{1}{\phi }-\frac{1-\psi }{1-\pi }\frac{\pi 
}{\psi }\left[ 1+\frac{1}{\phi }\right] =0.  \label{lowSS}
\end{equation}
The unique solution to \eqref{lowSS} is $\pi =\psi $ is the unique solution to
this equation. Since $p\left( \pi \right) =p\left( \psi \right) =0$ it
follows that $p\left( \psi \right) A_{q}\left( \psi \right) =0<Q^{\ast }$ so
this is an equilibrium as $\psi \leq \overline{\pi }.$ 
\end{proof}

\subsection{Proof of Proposition~\ref{higheq}}

\begin{proof}
We note that $\widetilde{G}$ is continuous, that $\widetilde{G}\left(0\right) >0,$ and $\widetilde{G}\left( \pi \right) \rightarrow -\infty $ as $\pi \rightarrow 1.$ Hence, the steady state condition (\ref{highSS}) has at least one solution $\pi ^{\ast }.$ From Lemma \ref{firmindifference} it is then immediate that $\left( \pi ^{\ast },p\left( \pi ^{\ast }\right) \right)
=\left( \pi ^{\ast },1\right) $ is inconsistent with equilibrium if $\pi
^{\ast }<\underline{\pi }.$ For $\pi ^{\ast }\in \left[ \underline{\pi },%
\overline{\pi }\right] $ is consistent with equilibrium if and only if
workers are willing to reject low tech offers, which is if and only if $%
A_{q}\left( \pi ^{\ast }\right) \geq Q^{\ast }.$ Finally, if $\pi ^{\ast }>%
\overline{\pi }$ it is direct from Lemma \ref{firmindifference} that only
high tech firms are willing to enter regardless of what workers do.%

\end{proof}

\subsection{Proof of Proposition~\ref{nonpicky}}

\begin{proof}

\begin{equation} \label{notpickySS0}
G\left( \overline{\pi },1,0\right) =\frac{\psi -\overline{\pi }}{\left( 1-\overline{\pi }\right) \psi }\left[ 1+\frac{1}{\phi }\right] >0
\end{equation}
if and only if $\psi >\overline{\pi }$, and
\begin{equation}\label{notpickySSp}
G\left( \overline{\pi },1,p\right) =G\left( \overline{\pi },1,0\right) 
+p\left[ \frac{A_{u}(\overline{\pi })}{\phi +(1-\phi )r}-\frac{1}{\phi }-\frac{1-\psi }{1-\overline{\pi }}\frac{\overline{\pi}}{\psi }\left( \frac{A_{q}(\overline{\pi })}{\phi }-\frac{1}{\phi }\right) \right].
\end{equation}

Hence, if $\psi >\overline{\pi }$ then $G\left( \overline{\pi },1,0\right)>0.$ So a unique solution to (\ref{notpickySS}) exists if $G\left( \overline{\pi },1,1\right) \leq 0,$ which is possible since
\begin{eqnarray}
\frac{A_{u}(\overline{\pi })}{\phi +(1-\phi )r}-\frac{1}{\phi }-\frac{1-\psi}{1-\overline{\pi }}\frac{\overline{\pi }}{\psi }\left( \frac{A_{q}(\overline{\pi })}{\phi }-\frac{1}{\phi }\right) 
< \frac{\psi-\overline{\pi}}{(1-\overline{\pi})\psi } \left( \frac{A_{q}(\overline{\pi })}{\phi} -\frac{1}{\phi} \right)
\leq 0. \notag 
\end{eqnarray}

If instead $\psi <\overline{\pi}$, then $G\left(\overline{\pi },1,p\right)$ is strictly increasing in $p\in \left[0,1 \right]$ since
\begin{eqnarray}
	\frac{A_{u}(\overline{\pi })}{\phi +(1-\phi )r}-\frac{1}{\phi }-\frac{1-\psi}{1-\overline{\pi }}\frac{\overline{\pi }}{\psi }\left( \frac{A_{q}(\overline{\pi })}{\phi }-\frac{1}{\phi }\right)
	> \frac{\psi-\overline{\pi}}{(1-\overline{\pi})\psi } \left[ \frac{A_{u}(\overline{\pi })}{\phi +(1-\phi )r}-\frac{1}{\phi}  \right] \notag
	> 0.
\end{eqnarray}
But
\begin{eqnarray}
	G\left(\overline{\pi},1,1\right)=1 + \frac{A_{u}(\overline{\pi })}{\phi +(1-\phi )r}-\frac{1-\psi}{1-\overline{\pi }}\frac{\overline{\pi }}{\psi }\left[ 1 + \frac{A_{q}(\overline{\pi })}{\phi } \right]
	< \frac{\psi-\overline{\pi}}{(1-\overline{\pi})\psi} \left[ 1 + \frac{A_{q}(\overline{\pi })}{\phi } \right]  \notag
	<0,
\end{eqnarray}
which implies that if $\psi <\overline{\pi}$ there is no solution $p\left(\overline{\pi}\right)\in \left[ 0,1 \right]$ to~\eqref{notpickySS} . \end{proof}

\subsection{Proof of Proposition~\ref{exist}}
\begin{proof}
	If $\psi \leq \overline{\pi }$ a low tech equilibrium exists. Hence,
	consider the case in which $\psi >\overline{\pi }.$ Assume first that $
	A_{q}\left( \overline{\pi }\right) \leq Q^{\ast }.$ Then, if $\widetilde{G}
	\left( \overline{\pi }\right) =G\left(\overline{\pi },0,1\right)=G\left( 
	\overline{\pi },1,1\right) \geq 0$ there exists a solution $\pi ^{\ast }\geq 
	\overline{\pi }$ to (\ref{highSS}) which corresponds to a high tech
	equilibrium (even if $\pi ^{\ast }=\overline{\pi }$ because $A_{q}\left( 
	\overline{\pi }\right) \leq Q^{\ast }$).\ $\ $Hence, for no high tech
	equilibrium to exist $\widetilde{G}\left( \overline{\pi }\right) =G\left( 
	\overline{\pi },0,1\right)=G\left( 
	\overline{\pi },1,1\right) < 0.$ Since we
	are also assuming $\psi >\underline{\pi }$ and $A_{q}\left( \overline{\pi }%
	\right) \leq Q^{\ast }$, Proposition \ref{nonpicky} guarantees existence of a
	pure strategy equilibrium at $\overline{\pi }.$ Hence, the only possibility
	for non-existence is that $A_{q}\left( \overline{\pi }\right) >Q^{\ast }$.
	Then, (assuming away the trivial case with $A_{q}\left( 0\right) \geq
	Q^{\ast }$) there exists $\Pi ^{\ast }<\overline{\pi }$ such that $%
	A_{q}\left( \Pi ^{\ast }\right) =Q^*.$ Assume that $\Pi ^{\ast }\leq 
	\underline{\pi }.$ Then for no high tech equilibrium to exist $\widetilde{G}%
	\left( \underline{\pi }\right) =G\left( \underline{\pi },0,1\right) <0$ as
	otherwise a solution $\pi ^{\ast }$ such that $\widetilde{G}\left( \pi
	^{\ast }\right) =0$ exists and corresponds to a high tech equilibrium$.$
	From (\ref{pos}) we know that $G\left( \underline{\pi },0,0\right) >0$ given
	that $\psi \geq \underline{\pi },$ so for no two sector equilibrium to exist
	at $\underline{\pi }$ it must be that $p\left( \underline{\pi }\right)
	A_{q}\left( \underline{\pi }\right) <Q^{\ast }$ for the unique solution to $%
	G\left( \underline{\pi },0,p\left( \underline{\pi }\right) \right) =0.$
	Since $G\left( \underline{\pi },0,p\right) $ in linear in $p$ and $p\left( 
	\underline{\pi }\right) <\frac{Q^{\ast }}{A_{q}\left( \underline{\pi }%
		\right) }$ it follows that $G\left( \underline{\pi },0,\frac{Q^{\ast }}{%
		A_{q}\left( \underline{\pi }\right) }\right) <0.$ Hence, for a mixed
	strategy equilibrium not to exist $G\left( \overline{\pi },1,\frac{Q^{\ast }%
	}{A_{q}\left( \overline{\pi }\right) }\right) \leq 0$ as solution is
	continuous in $\pi $ on $\left[ \underline{\pi },\overline{\pi }\right] .$
	We also have that $G\left( \overline{\pi },1,0\right) >0$ be because $\psi >%
	\underline{\pi },$ so, by linearity, if there is no fully mixed equilibrium
	there exists $p\left( \overline{\pi }\right) \in \left( 0,\frac{Q^{\ast }}{%
		A_{q}\left( \overline{\pi }\right) }\right] $ such that $G\left( \overline{%
		\pi },1,p\left( \overline{\pi }\right) \right) =0.$ Hence, we have
	established existence of at least one equilibrium whenever $\Pi ^{\ast }\leq 
	\underline{\pi }.$ The final possibility is that $\underline{\pi }<\Pi
	^{\ast }<\overline{\pi }.$ Then, for no high tech equilibrium to exist $%
	\widetilde{G}\left( \Pi ^{\ast }\right) =G\left( \Pi ^{\ast },\alpha \left(
	\Pi ^{\ast }\right) ,1\right) <0.$ For the same reason as above, for no
	mixed equilibrium to exist it must be that $G\left( \overline{\pi },1,\frac{%
		Q^{\ast }}{A_{q}\left( \overline{\pi }\right) }\right) \leq 0.$ Repeating
	the argument above it follows that there exists an equilibrium at $\overline{%
		\pi }.$
\end{proof}

\subsection{Proof of Proposition~\ref{discprop}}

\begin{proof}
Consider an alternative candidate equilibrium in which $\alpha ^{f}\left( 
\boldsymbol{\pi }\right) =1$ and $\alpha ^{m}\left( \boldsymbol{\pi }\right)
=0.$ In such an equilibrium, the following conditions must hold
\begin{eqnarray}
G\left(\pi^{f},1,p\left( \boldsymbol{\pi }\right) \right)&=&1+\frac{p\left( \boldsymbol{\pi }\right) A_{u}(\pi^{f})}{\phi +(1-\phi )r}+\frac{1-p\left( \boldsymbol{\pi }\right) }{\phi} -\frac{1-\psi }{1-\pi ^{f}}\frac{\pi^{f}}{\psi}\left[1+\frac{p\left( \boldsymbol{\pi }\right) A_{q}(\pi^{f})}{\phi }+\frac{1-p\left( \boldsymbol{\pi }\right) }{\phi }\right]=0, \notag \\
p\left( \boldsymbol{\pi }\right) A_{q}(\pi ^{f}) &\leq& Q^{\ast }, \notag \\
G\left( \pi ^{m},0,p\left( \boldsymbol{\pi }\right) \right)&=&1+\frac{ p\left( \boldsymbol{\pi }\right) A_{u}(\pi ^{m})}{\phi +(1-\phi )r}+\frac{1-p\left( \boldsymbol{\pi }\right) }{\phi } -\frac{1-\psi }{1-\pi ^{m}}\frac{\pi ^{m}}{\psi }\left[ 1+\frac{p\left( \boldsymbol{\pi }\right) A_{q}(\pi^{m})}{\phi }\right] =0, \notag \\
p\left(\boldsymbol{\pi }\right) A_{q}(\pi ^{m}) &\geq& Q^{\ast }.  \notag
\end{eqnarray} 
Note that 
\begin{eqnarray}
G\left( \pi ^{\ast },1,\frac{Q^{\ast }}{A_{q}\left( \pi ^{\ast }\right) }\right) -G\left( \pi ^{\ast },\alpha \left( \pi ^{\ast }\right), \frac{Q^{\ast }}{A_{q}\left( \pi ^{\ast }\right) }\right)
=\left( \alpha \left(\pi ^{\ast }\right) -1\right) \frac{1-\psi }{1-\pi ^{\ast }}\frac{\pi ^{\ast
}}{\psi }\left[ \frac{ 1-\frac{Q^{\ast }}{A_{q}\left( \pi ^{\ast }\right) } }{\phi }\right] <0. \notag
\end{eqnarray}

Since, $G\left( \pi ^{\ast },\alpha \left( \pi ^{\ast }\right), \frac{ Q^{\ast }}{A_{q}\left( \pi ^{\ast }\right) }\right) =0$ it follows that
$G\left( \pi ^{\ast },1,\frac{Q^{\ast }}{A_{q}\left( \pi ^{\ast }\right) } \right)<0.$ Symmetrically, 
\begin{eqnarray}
G\left( \pi ^{\ast },0,\frac{Q^{\ast }}{A_{q}\left( \pi ^{\ast }\right) } \right) -G\left( \pi ^{\ast },\alpha \left( \pi ^{\ast }\right), \frac{ Q^{\ast }}{A_{q}\left( \pi ^{\ast }\right) }\right) 
=\alpha \left( \pi^{\ast }\right) \frac{1-\psi }{1-\pi ^{\ast }}\frac{\pi ^{\ast }}{\psi}\left[ \frac{ 1-\frac{Q^{\ast }}{A_{q}\left( \pi ^{\ast }\right) } }{\phi}
\right] >0, \notag
\end{eqnarray}
so $G\left( \pi ^{\ast },0,\frac{Q^{\ast }}{A_{q}\left( \pi ^{\ast }\right) } \right) >0$.

Notice that $G\left(0,1,\frac{Q^{\ast}}{A_{q}\left(\pi^{\ast}\right)}\right)>0 $ and $ \lim_{\pi \rightarrow 1}G\left(\pi ,0,\frac{Q^{\ast }}{A_{q}\left( \pi ^{\ast }\right) }\right) 
= -\infty $. Hence,
there exists $\left(\pi^f,\pi^m\right)$ with $\pi^f<\pi^{\ast}<\pi ^m$ such that
\begin{eqnarray}
G\left( \pi ^{f},1,\frac{Q^{\ast }}{A_{q}\left( \pi ^{\ast }\right) }\right)
&=&0 \\
G\left( \pi ^{m},0,\frac{Q^{\ast }}{A_{q}\left( \pi ^{\ast }\right) }\right)
&=&0.  \notag
\end{eqnarray}
It is thus immediate that $A_{q}\left( \pi ^{f}\right) \frac{Q^{\ast }}{%
	A_{q}\left( \pi ^{\ast }\right) }<Q^{\ast }$ and $A_{q}\left( \pi
^{m}\right) \frac{Q^{\ast }}{A_{q}\left( \pi ^{\ast }\right) }>Q^{\ast },$
justifying the incentives to accept and reject low tech jobs for the two groups. In contrast, the indifference condition for the entrant firms is not necessarily satisfied. However,%
\begin{eqnarray*}
	1 &>&\pi ^{\ast }\alpha \left( \pi ^{\ast }\right) +\left( 1-\pi ^{\ast
	}\right)  \\
	&=&\pi ^{\ast }A_{q}(\pi ^{\ast })W_{q}+(1-\pi ^{\ast })A_{u}(\pi ^{\ast
	})W_{u} \\
	&>&\pi ^{f}A_{q}(\pi ^{f})W_{q}+(1-\pi ^{f})A_{u}(\pi ^{f})W_{u},
\end{eqnarray*}
and
\begin{eqnarray*}
	\left( 1-\pi ^{m}\right)  &<&1-\pi ^{\ast }<\pi ^{\ast }\alpha \left( \pi
	^{\ast }\right) +\left( 1-\pi ^{\ast }\right)\\
	&=&\pi ^{\ast }A_{q}(\pi ^{\ast })W_{q}+(1-\pi ^{\ast })A_{u}(\pi ^{\ast
	})W_{u} \\
	&<&\pi ^{m}A_{q}(\pi ^{m})W_{q}+(1-\pi ^{m})A_{u}(\pi ^{m})W_{u},
\end{eqnarray*}
so there exists a unique $\left( \lambda ^{f},\lambda ^{m}\right) $ with $\lambda ^{f}+\lambda ^{m}=1$ such that%
\begin{eqnarray*}
	\left[ \lambda ^{f}+\lambda ^{m}\left( 1-\pi ^{m}\right) \right] W_{l}
	&=&\lambda ^{f}\left[ \pi ^{f}A_{q}(\pi ^{f})W_{q}+(1-\pi ^{f})A_{u}(\pi
	^{f})W_{u}\right]  \\
	&+&\lambda ^{m}\left[ \pi ^{m}A_{q}(\pi ^{m})W_{q}+(1-\pi ^{m})A_{u}(\pi
	^{m})W_{u}\right].
\end{eqnarray*}
By continuity of $G$ there is an interval $\left( \underline{p},\overline{p}
\right) $ around $\frac{Q^{\ast }}{A_{q}\left( \pi ^{\ast }\right) }$ such
that $pA_{q}(\pi ^{f})<Q^{\ast }$ and $pA_{q}(\pi ^{m})>Q^{\ast }$ and  
\begin{eqnarray*}
	G\left( 0,1,p\right)>&0&>G\left( \pi ^{\ast },1,p\right)  \\
	G\left( \pi ^{\ast },1,p\right)>&0&>\lim_{\pi \rightarrow 1}G\left( \pi
	,1,p\right) 
\end{eqnarray*}
for each $p\in \left( \underline{p},\overline{p}\right)$. For each $p$ in the interval there is a corresponding solution $\pi ^{f}\left( p\right) ,\pi
^{m}\left( p\right) $ such that $\pi ^{f}\left( p\right) <\pi ^{\ast }<\pi
^{m}\left( p\right) $ and for each such solution there is a unique $\left(
\lambda ^{f}\left( p\right) ,\lambda ^{m}\left( p\right) \right) $ with $%
\lambda ^{f}\left( p\right) +\lambda ^{m}\left( p\right) =1$ that makes firms
indifferent across sectors. Moreover, if we pick $\left( \pi ^{f}\left(
p\right) ,\pi ^{m}\left( p\right) \right) $ as always being the smallest
solutions differentiability of $G\left( \pi ,\alpha ,p\right) $ on $\left(
0,1\right) $ implies that within the interval 
\begin{eqnarray}
\frac{\partial G\left( \pi ^{f}\left( p\right) ,1,p\right) }{\partial \pi
	^{f}} &<&0, \\
\frac{\partial G\left( \pi ^{m}\left( p\right) ,0,p\right) }{\partial \pi
	^{f}} &<&0. \notag
\end{eqnarray}%
By a direct calculation%
\begin{equation*}
\frac{\partial G\left( \pi ,\alpha ,p\right) }{\partial p}=\frac{A_{u}(\pi )%
}{\phi +(1-\phi )r}-\frac{1}{\phi }-\frac{1-\psi }{1-\pi }\frac{\pi }{\psi }%
\left[ 1+\frac{A_{q}(\pi )}{\phi }-\frac{\alpha }{\phi }\right].
\end{equation*}
Hence, $\frac{\partial G\left( \pi ^{m}\left( p\right) ,0,p\right) }{\partial p}<0$ whereas $\frac{\partial G\left( \pi ^{f}\left( p\right),0,p\right) }{\partial p}$ is ambiguous.
However, if $\phi $ is large enough
we can make sure that $\frac{\partial G\left( \pi ^{f}\left( p\right),1,p\right) }{\partial p}\leq 0$ for $p\in \left( \underline{p},\overline{p}%
\right)$ in which case a standard implicit differentiation implies that $\pi ^{f}\left( p\right)$ and $\pi ^{m}\left( p\right) $ are both decreasing in $p$. It then follows that for firms to keep indifference it is necessary
that $\lambda ^{m}\left( p\right) $ is strictly increasing in $p$.

\end{proof}

\section{Appendix B: Numerical Examples} \label{appexamples}

\subsection{Description of Function~\eqref{Gdef}}
We describe the G function (\ref{Gdef}) piecewisely as follows. For $\pi<\underline{\pi}$,~\eqref{Gdef} evaluated at $p(\pi)=0$ and $\alpha(\pi)=1$ 
\begin{equation}\label{Glow}
G\left(\pi,1,0 \right) = 1+\frac{1}{\phi} -\frac{1-\psi}{1-\pi} \frac{\pi}{\psi} \left[1 +\frac{1}{\phi} \right].
\end{equation}
For $\pi>\overline{\pi}$, $p(\pi)=1$, $\alpha(\pi)$ is indetermined,~\eqref{Gdef} evaluated at $p(\pi)=1$
\begin{equation}\label{Ghigh}
G\left(\pi,\alpha(\pi),1 \right) = 1+\frac{A_u(\pi)}{\phi+(1-\phi)r} -\frac{1-\psi}{1-\pi} \frac{\pi}{\psi} \left[1 +\frac{A_q(\pi)}{\phi} \right].
\end{equation}
For $\pi\in (\underline{\pi},\overline{\pi})$, $\alpha(\pi) \in (0,1)$, $p(\pi)\in [0,1]$, (\ref{Gdef}) evaluated at $p(\pi)A_q(\pi)= Q^*$
\begin{eqnarray}\label{Ghl}
G\left( \pi, \alpha(\pi),\frac{Q^*}{A_q(\pi)}  \right)
= 1+\frac{ \frac{Q^*}{A_q(\pi)}  A_u(\pi)}{\phi+(1-\phi)r} + \frac{1- \frac{Q^*}{A_q(\pi)}  }{\phi}
-\frac{1-\psi}{1-\pi} \frac{\pi}{\psi} 
\Bigg[ 1 +\frac{\frac{Q^*}{A_q(\pi)}  A_q(\pi)}{\phi}+\frac{ 1-\frac{Q^*}{A_q(\pi)}  }{\phi} \alpha(\pi) \Bigg],
\end{eqnarray}
where 
\begin{equation*}
   \alpha(\pi) = \frac{\pi A_{q}(\pi)W_{q}+(1-\pi )A_{u}(\pi )W_{u}-(1-\pi)W_l }{\pi W_l }.
\end{equation*}
At $\pi = \underline{\pi}$, $\alpha(\underline{\pi})=0$, \eqref{Gdef} evaluated at $\underline{\pi}$ and $\alpha(\underline{\pi})=0$
\begin{equation}\label{Ghl_l}
G\left(\underline{\pi},0,p(\underline{\pi}) \right) =1+\frac{p(\underline{\pi}) A_u(\underline{\pi})}{\phi+(1-\phi)r}+\frac{1-p(\underline{\pi})}{\phi}
-\frac{1-\psi}{1-\underline{\pi}} \frac{\underline{\pi} }{\psi} \left[1 +\frac{p(\underline{\pi}) A_q(\underline{\pi})}{\phi} \right].
\end{equation}
At $\pi = \overline{\pi}$, $\alpha(\overline{\pi})=1$,~\eqref{Gdef} evaluated at $\overline{\pi}$ and $\alpha(\overline{\pi})=1$
\begin{eqnarray}
G\left(\overline{\pi},1,p(\overline{\pi}) \right) =1+\frac{p(\overline{\pi})A_u(\overline{\pi})}{\phi+(1-\phi)r}+\frac{1-p(\overline{\pi})}{\phi}
 -\frac{1-\psi}{1-\overline{\pi}} \frac{\overline{\pi} }{\psi} \left[1 +\frac{p(\overline{\pi})  A_q(\overline{\pi})}{\phi}+\frac{1-p(\overline{\pi}) }{\phi} \right]. \label{Ghl_h}
\end{eqnarray}

In sum, we simplify function~\eqref{Gdef} into a function of one dimension. For $\pi\in \left[ 0,1 \right]$ except $\overline{\pi}$ and $\underline{\pi}$,~\eqref{Gdef} is a function of $\pi$
\begin{equation*}
G_0(\pi)=
\begin{cases}
1+\frac{1}{\phi} -\frac{1-\psi}{1-\pi} \frac{\pi}{\psi} \left[1 +\frac{1}{\phi} \right] &\text{if }\pi\in[0,\underline{\pi}) \\
1+\frac{ \frac{Q^*}{A_q(\pi)} A_u(\pi) }{\phi+(1-\phi)r}+\frac{1- \frac{Q^*}{A_q(\pi)} }{\phi} 
-\frac{1-\psi}{1-\pi} \frac{\pi}{\psi} \left[1 +\frac{ \frac{Q^*}{A_q(\pi)} A_q(\pi)}{\phi} \right. \\
\left.\hspace{1.5in} +\frac{1-\frac{Q^*}{A_q(\pi)} }{\phi} 
\frac{\pi A_{q}(\pi)W_{q}+(1-\pi ) A_{u}(\pi )W_{u}-(1-\pi)W_l }{\pi W_l } \right] &\text{if }\pi\in(\underline{\pi},\overline{\pi} ). \notag \\
1+\frac{A_u(\pi)}{\phi+(1-\phi)r} -\frac{1-\psi}{1-\pi} \frac{\pi}{\psi} \left[1 +\frac{A_q(\pi)}{\phi} \right] &\text{if }\pi \in (\overline{\pi},1] \notag
\end{cases}
\end{equation*}

At $\overline{\pi}$ and $\underline{\pi}$,~\eqref{Gdef} is a function of $p$
\begin{equation*}
G_1(p)=
\begin{cases}
1+\frac{p(\underline{\pi}) A_u(\underline{\pi})}{\phi+(1-\phi)r}+\frac{1-p(\underline{\pi}) }{\phi}
-\frac{1-\psi}{1-\underline{\pi}} \frac{\underline{\pi} }{\psi} \left[1 +\frac{p(\underline{\pi}) A_q(\underline{\pi})}{\phi} \right] &\text{if $\pi=\underline{\pi} $} \notag \\
1+\frac{p(\overline{\pi}) A_u(\overline{\pi})}{\phi+(1-\phi)r}+\frac{1-p(\overline{\pi})}{\phi}-\frac{1-\psi}{1-\overline{\pi}} \frac{\overline{\pi} }{\psi} \left[1 +\frac{p(\overline{\pi}) A_q(\overline{\pi})}{\phi}+\frac{1-p(\overline{\pi}) }{\phi} \right] &\text{if $\pi=\overline{\pi} $}.  \notag 
\end{cases}
\end{equation*}	

In all the numerical examples, we use the signal distributions $f_q(\theta)=2\theta$, $f_u(\theta)=2(1-\theta)$, which satisfy the property of monotone likelihood ratio. Calibrating $W_q=1$ and $W_u=-1$. Then the hiring probability for qualified and unqualified workers are $A_q(\pi)=\pi(2-\pi)$ and $A_u(\pi)=\pi^2$, respectively.
\begin{proof}
	With $F_q(\theta)=\theta^2$, $F_u(\theta)=2\theta-\theta^2$, firms' hiring threshold from problem~\ref{maxproblem} is
	\begin{equation}
	s(\pi)=\arg \max_{\theta }\pi \left[ 1- \theta^2 \right] - (1-\pi )\left[1-2\theta+\theta^2 \right],
	\end{equation}
    so $s(\pi)=1-\pi$, and
	\begin{equation*}
		A_q(\pi) =1-F_q(s(\pi)) = 1-(1-\pi)^2=2\pi-\pi^2, \quad A_u(\pi)=1-F_u(s(\pi))= \pi^2.
	\end{equation*}
\end{proof} 

All the graphs are plotted by plugging $A_q(\pi)$ and $A_u(\pi)$ into the function~\eqref{Gdef} under different calibrated parameters.

\pagebreak
\bibliography{references}

\pagebreak

\section{External Appendix: Not Intended For Publication}

\subsection{Other Examples in Baseline Model}\label{othersymmetric}
There exist other unique equilibria and multiple equilibria in the baseline model. We keep the same parameters for $W_q=1$, $W_u=-1$, $y_l=0.5$, $w_l=0.495$, while changing $\beta$, $\phi$, $r$, $\psi$ and $b$.

\bigskip
\noindent \textbf{Other unique equilibria examples}.

\begin{center}
	\begin{tabular}{ c|c|c|c|c|c|c } 
		\hline
		Unique Equilibria  & $\beta$ & $\phi$ & $r$ & $\psi$ & $b$ & In Equilibrium \\ 
		\hline
		Example (a) & .99 & .15 & .75 & .075 & .2 & only low tech \\ 
		Example (b) & .99 & .15 & .75 & .75 &  .2 & only high tech\\
		Example (c) & .9 & .06 & .75  & .25 &  .2 & two sectors with low tech jobs rejected\\ 
		Example (d) & .8 & .06 & .75 &  .25 &  .2 & two sectors with low tech job accepted\\
		\hline
	\end{tabular}
\end{center}

\begin{itemize}
	\item[(a)] With $\beta=.99, \phi=.15, r=.75, \psi=.075, b=.2$, there is a unique equilibria in which only low tech sector is active. An important change is that the mass of skilled workers is decreased relative to the other examples. From Figure~\ref{unique_lowtech} we see that there exists a low tech equilibria, and there is neither a mixed, nor a high tech equilibrium.
	
	\begin{figure}[H]
		\centering
		\includegraphics[width=.5\linewidth]{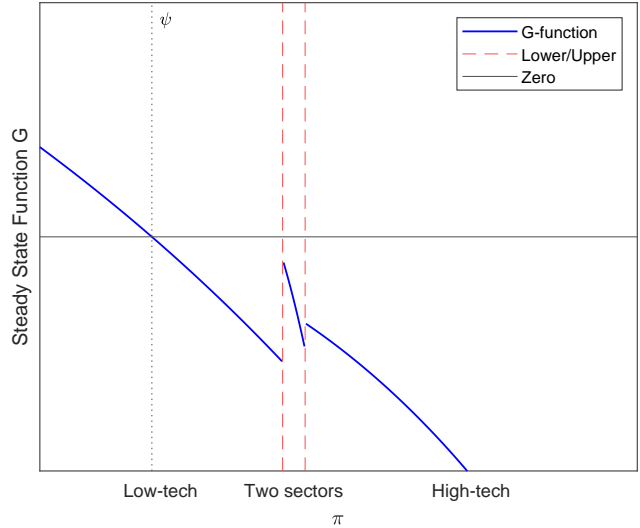}
		\caption{$\beta = 0.99$, $\phi = 0.15$, $r = 0.75$, $\psi = 0.075$, $b = 0.2$.}
		\label{unique_lowtech}
	\end{figure}
	At $\pi=\underline{\pi}$, solving for the steady state value $p(\underline{\pi}) = 0.04626 \in \left[ 0,1 \right]$, but the worker's optimality for $\alpha(\underline{\pi})=0$ does not hold because $p(\underline{\pi})A_q(\underline{\pi})-Q^*= -0.0504<0$. At $\pi=\overline{\pi}$, the steady state value $p(\overline{\pi})=2.3693\notin\left[0,1 \right]$.
	
	\item[(b)] With $\beta=.99, \phi=.15, r=.75, \psi=.75, b=.2$, there is a unique equilibria in which only high tech sector is active. From Figure~\ref{unique_hightech} we see that there exists a high tech equilibria, and there is neither a mixed, nor a low tech equilibrium.
	
	\begin{figure}[H]
		\centering
		\includegraphics[width=.5\linewidth]{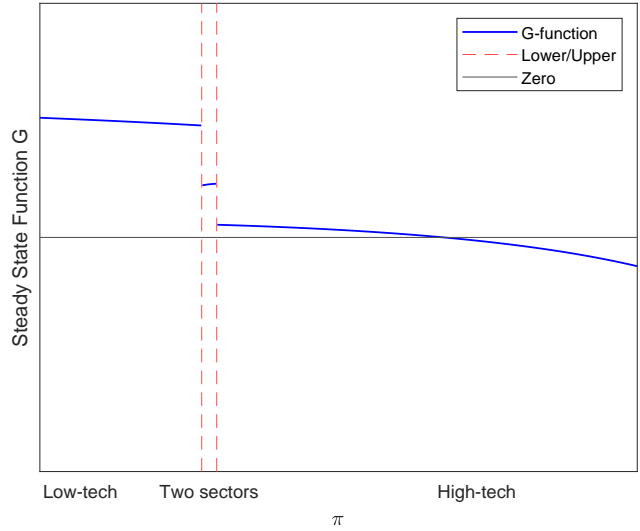}
		\caption{$\beta = 0.99$, $\phi = 0.15$, $r = 0.75$, $\psi = 0.75$, $b = 0.2$.}
		\label{unique_hightech}
	\end{figure}
	At $\pi=\underline{\pi}$, the steady state value $p(\underline{\pi}) = 1.1242 \notin \left[ 0,1 \right]$. At $\pi=\overline{\pi}$, the steady state value $p(\overline{\pi})=1.1290\notin\left[0,1 \right]$.
	
	\item[(c)] With $\beta=.9, \phi=.06, r=.75, \psi=.25, b=.2$, there is a unique equilibria in which both sectors are active and the low tech jobs are rejected (accepting is not an equilibrium). This is the case of uniqueness in the baseline model in Section~\ref{exunique}.  
	
	\item[(d)] With $\beta=.8, \phi=.06, r=.75, \psi=.25, b=.2$, there is a unique equilibria in which both sectors are active and the low tech jobs are accepted (rejecting is not an equilibria). From Figure~\ref{unique_twosectors1} we see that there is neither a mixed, nor a high tech nor a low tech equilibrium.
	
	\begin{figure}[H]
		\centering
		\includegraphics[width=.5\linewidth]{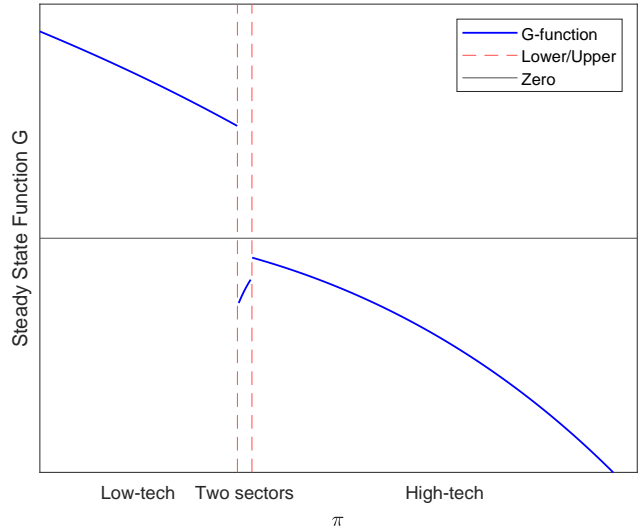}
		\caption{$\beta = 0.8$, $\phi = 0.06$, $r = 0.75$, $\psi = 0.25$, $b = 0.2$.}
		\label{unique_twosectors1}
	\end{figure}
	Solving for the unique $p(\underline{\pi})\in \left[0,1\right]$ that is consistent with steady state we find that under the parameter of the example $p(\underline{\pi})A_q(\underline{\pi}) - Q^*= -0.0774<0$, so rejecting low tech offers is inconsistent with worker optimality. At $\pi=\overline{\pi}$, the steady state value $p(\overline{\pi})=0.8433\in\left[0,1 \right]$ is consistent with workers' accepting low tech offers as $p(\overline{\pi})A_q(\overline{\pi})- Q^*=-0.0844<0$.
	
\end{itemize}

\bigskip
\noindent \textbf{Other Multiple Equilibria Examples} in the baseline model.

\begin{center}
	\begin{tabular}{ c|c|c|c|c|c|p{5cm} } 
		\hline
		Multiple Equilibria & $\beta$ & $\phi$ & $r$ & $\psi$ & $b$ & \\ 
		\hline
		Two equilibria    & .99 & .06 & .75 (.6) & .25 & .2 & (1) two sectors with low tech jobs rejected; (2) two sectors with workers mixing \\
		\hline
		Three equilibria & .99 & .08 & .75 & .25 & .2 &  (1) two sectors with low tech jobs accepted; (2) two sectors with workers mixing; (3) two sectors with low tech jobs rejected \\ 
		\hline
		Three equilibria  & .99 & .06 & .75 & .2 & .2 &  (1) low tech; (2) two sectors with workers mixing; (3) two sectors with low tech jobs rejected \\ 
		\hline
	\end{tabular}
\end{center}

\subsection{Single Sector Equilibria with Groups}\label{singlewithgroups}
\textbf{Only low tech firms.} In the case of only low tech firms $p(\boldsymbol{\pi})=0$, so $\alpha^m(\boldsymbol{\pi})=\alpha^f(\boldsymbol{\pi})=1$, the steady state functions for $j=f,m$ are
\begin{equation*}
G(\pi^j,1,0) = 1+\frac{1}{\phi } - \frac{1-\psi }{1-\pi ^{j}} \frac{\pi ^{j}}{\psi }\left[ 1+\frac{1}{\phi }\right].
\end{equation*} 
So $\pi^f=\pi^m=\psi$. 

Moreover, firms' entry optimality for $p(\boldsymbol{\pi})=0$ requires
\begin{equation*}
\sum_{j=f,m} \lambda^j \left[ \pi ^{j}\alpha ^{j}\left( \boldsymbol{\pi }\right) +(1-\pi^{j})\right] W_{l}
\geq \sum_{j=f,m} \lambda^j \left[ \pi ^{j}A_{q}(\pi ^{j})W_{q}+(1-\pi
^{j})A_{u}(\pi ^{j})W_{u}\right],
\end{equation*}
degenerating into the baseline model
\begin{equation*}
W_{l} \geq \left[ \psi A_{q}(\psi )W_{q}+(1-\psi)A_{u}(\psi)W_{u}\right],
\end{equation*}
so $p(\boldsymbol{\pi})=0$ requires $\psi\leq \overline{\pi}$.

\begin{lemma}
	There is a symmetric equilibrium in which only low tech firms are active and $\pi^m=\pi^f=\psi$ if and only if $\psi\leq \overline{\pi}$.
\end{lemma}

\noindent \textbf{Only high-tech firms.} In the case of only high-tech firms $p(\boldsymbol{\pi})=1$, so workers' optimality of acceptance rule doesn't matter. The steady state functions for $j=f,m$ are
\begin{eqnarray}
G(\pi^f,\alpha^f,p(\boldsymbol{\pi})=1 ) &=&1+\frac{A_u(\pi^f)}{\phi+(1-\phi)r} -\frac{1-\psi}{1-\pi^f} \frac{\pi^f}{\psi} \left[1 +\frac{A_q(\pi^f)}{\phi} \right] = 0, \notag \\
G(\pi^m,\alpha^m,p(\boldsymbol{\pi})=1 ) &=&1+\frac{A_u(\pi^m)}{\phi+(1-\phi)r} -\frac{1-\psi}{1-\pi^m} \frac{\pi^m}{\psi} \left[1 +\frac{A_q(\pi^m)}{\phi} \right] = 0, \notag 
\end{eqnarray}
which pin down $\pi^m$ and $\pi^f$. Moreover, firms' entry optimality for $p(\boldsymbol{\pi})=1$ requires
\begin{equation*}
\sum_{j=f,m} \lambda^j \left[ \pi ^{j}\alpha ^{j}\left( \boldsymbol{\pi }\right) +(1-\pi^{j})\right] W_{l}
\leq \sum_{j=f,m} \lambda^j \left[ \pi ^{j}A_{q}(\pi ^{j})W_{q}+(1-\pi
^{j})A_{u}(\pi ^{j})W_{u}\right],
\end{equation*}
which is easily valid because $\alpha^f$ and $\alpha^m$ are free if $p(\boldsymbol{\pi})=0$.

From the baseline analysis, we know $G\left(\pi,\alpha,p(\boldsymbol{\pi})=1\right)=1+\frac{A_u(\pi)}{\phi+(1-\phi)r} -\frac{1-\psi}{1-\pi} \frac{\pi}{\psi} \left[1 +\frac{A_q(\pi)}{\phi} \right]=0$ has at least one solution.
\begin{lemma}
	If $G\left(\pi,\alpha,p(\boldsymbol{\pi})=1\right) = 0$ has a unique solution, then there is a unique symmetric equilibrium in which only high tech firms are active and $\pi^f=\pi^m$. Otherwise, either $\pi^f=\pi^m$ or $\pi^f\neq \pi^m$.
\end{lemma}

\subsection{Other Asymmetric Equilibria}\label{otherasymmetric}
Similarly, we can construct equilibria in which $\pi^m>\pi^f$, $\alpha^m\in(0,1)$, $\alpha^f=1$ under some parameters.

\bigskip
\noindent \textbf{Only group $m$ is indifferent}: If $\alpha^m(\boldsymbol{\pi})\in(0,1)$, then $\alpha^f(\boldsymbol{\pi})= 1$. To make $\{ \pi^m>\pi^f,\alpha^m\in(0,1),\alpha^f=1 \}$ an equilibrium, $\pi^f$ and $\pi^m$ should satisfy:
\begin{itemize}
	\item[(1)] Firms' indifference condition:
	\begin{equation*}
	\lambda^m \left[ \pi ^{m}\alpha ^{m}\left( \boldsymbol{\pi }\right) +(1-\pi^{m})\right] W_{l}+\lambda^f W_{l}
	=\sum_{j=f,m} \lambda^j \left[ \pi ^{j}A_{q}(\pi ^{j})W_{q}+(1-\pi
	^{j})A_{u}(\pi ^{j})W_{u}\right],
	\end{equation*}
	$\alpha ^{m}\left( \boldsymbol{\pi }\right)$ is increasing in both $\pi^f$ and $\pi^m$.
	\item[(2)] Steady state conditions: $G(\pi^m,\alpha^m(\boldsymbol{\pi}),p(\boldsymbol{\pi}))=0$, $G(\pi^f,\alpha^f(\boldsymbol{\pi})=1,p(\boldsymbol{\pi}))=0$
	\begin{eqnarray}
	1+\frac{p\left( \boldsymbol{\pi }\right) A_{u}(\pi ^m)}{\phi +(1-\phi )r}+\frac{1-p\left( \boldsymbol{\pi }\right) }{\phi }&=&\frac{1-\psi }{1-\pi ^m }\frac{\pi ^m }{\psi }\left[ 1+\frac{p\left( \boldsymbol{\pi }\right) A_{q}(\pi^m )}{\phi }+\frac{(1-p\left( \boldsymbol{\pi }\right) )\alpha^m \left(\boldsymbol{\pi }\right) }{\phi }\right] \notag \\
	1+\frac{p\left( \boldsymbol{\pi }\right) A_{u}(\pi ^f )}{\phi +(1-\phi )r}+\frac{1-p\left( \boldsymbol{\pi }\right) }{\phi }&=&\frac{1-\psi }{1-\pi ^f }\frac{\pi ^f }{\psi }\left[ 1+\frac{p\left( \boldsymbol{\pi }\right) A_{q}(\pi^f )}{\phi }+\frac{ 1-p\left( \boldsymbol{\pi }\right)  }{\phi }\right] \notag
	\end{eqnarray}
	\item[(3)] Group $m$ workers' indifference: $Q^* = p(\boldsymbol{\pi}) A_q(\pi^m)$
\end{itemize}
	Then, $\{\pi^m,\pi^f,\alpha^m(\boldsymbol{\pi}),p(\boldsymbol{\pi }) \}$ are pinned down by the four equations. Substituting out $p(\boldsymbol{\pi})$ and $\alpha^m(\boldsymbol{\pi})$, the equilibrium $\pi^m$, $\pi^f$ are determined by
\begin{eqnarray}
&&1+ \frac{Q^*}{A_q(\pi^m)} \frac{A_{u}(\pi ^m)}{\phi +(1-\phi )r}+\frac{1- \frac{Q^*}{A_q(\pi^m)} }{\phi } = \frac{1-\psi }{1-\pi ^m }\frac{\pi ^m }{\psi }\left[ 1+ \frac{Q^*}{\phi }+\frac{1- \frac{Q^*}{A_q(\pi^m)} }{\phi } \right. \notag \\
&& \hspace{2cm} \left. \frac{\sum_{j=f,m} \lambda^j \left[ \pi ^{j}A_{q}(\pi ^{j})W_{q}+(1-\pi^{j})A_{u}(\pi ^{j})W_{u}\right] - \lambda^f W_l - \lambda^m (1-\pi^m) W_l}{\lambda^m \pi^{m} W_{l} } \right], \notag \\
&&1+ \frac{A_{u}(\pi ^f ) }{A_q(\pi^m)} \frac{Q^*}{\phi +(1-\phi )r}+\frac{1- \frac{Q^*}{A_q(\pi^m)} }{\phi } = \frac{1-\psi }{1-\pi ^f }\frac{\pi ^f }{\psi }\left[ 1+ \frac{Q^*}{A_q(\pi^m)} \frac{A_{q}(\pi^f )}{\phi }+\frac{ 1- \frac{Q^*}{A_q(\pi^m)} }{\phi } \right]. \notag
\end{eqnarray}
Lastly, check the group $f$ workers' conditions: $p(\boldsymbol{\pi})A_q(\pi^f)< Q^*$. Let $m=\min p(\boldsymbol{\pi})A_q(\pi^f)$ subject to $\pi^f,\pi^m$ are solutions to the above two equations. Then if $m<Q^*$ there exists some equilibrium in which group $m$ workers are indifferent and group $f$ workers reject the low tech job offers.

\bigskip
\noindent \textbf{Neither group is indifferent.} There are four cases: $\{\alpha^m=0,\alpha^f=1 \}$, $\{\alpha^m=1,\alpha^f=0 \}$, $\{\alpha^m=1,\alpha^f=1 \}$, $\{\alpha^m=0,\alpha^f=0 \}$. 

With $\pi^m>\pi^f$, the only interesting one is: $\{\alpha^m=0, \alpha^f=1 \}$. To make $\{ \pi^m > \pi^f,\alpha^m=0,\alpha^f=1 \}$ be an equilibrium, it should satisfy
\begin{itemize}
	\item[(1)] Firms' indifference: 
	\begin{equation*}
	\lambda^m (1-\pi^{m}) W_{l}+\lambda^f W_{l}
	=\sum_{j=f,m} \lambda^j \left[ \pi ^{j}A_{q}(\pi ^{j})W_{q}+(1-\pi
	^{j})A_{u}(\pi ^{j})W_{u}\right]
	\end{equation*}
	\item[(2)] Steady state condition: $G(\pi^m,\alpha^m(\boldsymbol{\pi})=0,p(\boldsymbol{\pi})))=0$, $G(\pi^f,\alpha^f(\boldsymbol{\pi})=1,p(\boldsymbol{\pi})))=0$
	\begin{eqnarray}
	1+\frac{p\left( \boldsymbol{\pi }\right) A_{u}(\pi ^m)}{\phi +(1-\phi )r}+\frac{1-p\left( \boldsymbol{\pi }\right) }{\phi }&=&\frac{1-\psi }{1-\pi ^m }\frac{\pi ^m }{\psi }\left[ 1+\frac{p\left( \boldsymbol{\pi }\right) A_{q}(\pi^m )}{\phi} \right] \notag \\
	1+\frac{p\left( \boldsymbol{\pi }\right) A_{u}(\pi ^f )}{\phi +(1-\phi )r}+\frac{1-p\left( \boldsymbol{\pi }\right) }{\phi }&=&\frac{1-\psi }{1-\pi ^f }\frac{\pi ^f }{\psi }\left[ 1+\frac{p\left( \boldsymbol{\pi }\right) A_{q}(\pi^f )}{\phi }+\frac{ 1-p\left( \boldsymbol{\pi }\right)  }{\phi }\right]. \notag 
	\end{eqnarray}
\end{itemize} 
	Then $\{\pi^m,\pi^f,p(\boldsymbol{\pi }) \}$ are pinned down by the three equations, and the workers' conditions hold: $p(\boldsymbol{\pi})A_q(\pi^f)\leq Q^*\leq p(\boldsymbol{\pi}) A_q(\pi^m)$.

\end{document}